\documentclass[preprint,showpacs]{revtex4}
\usepackage{epsfig}
\newcommand{\ben}{\begin{eqnarray}}
\newcommand{\een}{\end{eqnarray}}

\newcommand{\bef}{\begin{figure}[h!bt]\centering}
\newcommand{\eef}{\end{figure}}
\newcommand{\bet}{\begin{table}[hbt]\centering}
\newcommand{\eet}{\end{table}}

\begin{document}
\title{Temperature-doping phase diagrams for ${\rm Ba(Fe}_{1-x}{\rm TM}_x)_2{\rm As}_2$(TM =Ni,
Cu, Cu / Co) single crystals}
\author{N. Ni}
\affiliation{Ames Laboratory and
Department of Physics and Astronomy, Iowa State University, Ames,
IA 50011, USA}

\author{A. Thaler}
\affiliation{Ames Laboratory and Department of Physics and
Astronomy, Iowa State University, Ames, IA 50011, USA}

\author{J. Q. Yan}
\affiliation{Ames Laboratory and
Department of Physics and Astronomy, Iowa State University, Ames,
IA 50011, USA}

\author{A. Kracher}
\affiliation{Ames Laboratory and
Department of Physics and Astronomy, Iowa State University, Ames,
IA 50011, USA}

\author{E. Colombier}
\affiliation{Ames Laboratory and Department of Physics and
Astronomy, Iowa State University, Ames, IA 50011, USA}
\author{S. T. Hannahs}
\affiliation{National High Magnetic Field Laboratory, 1800 East Paul Dirac Drive, Tallahassee, Florida 32310, USA}

\author{S. L. Bud'ko}
\affiliation{Ames Laboratory and Department of Physics and
Astronomy, Iowa State University, Ames, IA 50011, USA}

\author{P. C. Canfield}
\affiliation{Ames Laboratory and Department of Physics and
Astronomy, Iowa State University, Ames, IA 50011, USA}

\begin{abstract}
Microscopic, structural, transport and thermodynamic
measurements of single crystalline ${\rm Ba(Fe}_{1-x}{\rm TM}_x)_2{\rm As}_2$ (TM =
Ni and Cu) series, as well as two mixed TM = Cu / Co series, are reported.
In addition, high magnetic field, anisotropic $H_{c2}(T)$ data were measured up to 33 T for the optimally Ni doped ${\rm BaFe}_2{\rm As}_2$ sample.
All the transport and thermodynamic measurements indicate that the
structural and magnetic phase transitions at 134 K in pure ${\rm BaFe}_2{\rm As}_2$ are monotonically suppressed and increasingly separated
in a similar manner by these dopants.
In the ${\rm Ba(Fe}_{1-x}{\rm Ni}_x)_2{\rm As}_2$ ($x \leq 0.072$), superconductivity, with $T_c$ up to 19 K,
is stabilized for $0.024\leq x \leq0.072$.
In the ${\rm Ba(Fe}_{1-x}{\rm Cu}_x)_2{\rm As}_2$ ($x\leq0.356$) series, although the structural and magnetic transitions are suppressed, there
is only a very limited region of superconductivity: a sharp drop of the resistivity to zero near 2.1 $K$ is found only for the $x=0.044$ samples.
In the ${\rm Ba(Fe}_{1-x-y}{\rm Co}_x{\rm Cu}_y)_2{\rm As}_2$ series, superconductivity, with $T_c$ values up to 12 $K$ ($x\sim 0.022$ series)
and 20 $K$ ($x\sim 0.047$ series), is stabilized.
Quantitative analysis of the detailed temperature-dopant concentration ($T-x$)
and temperature-extra electrons ($T-e$) phase diagrams of these series shows that there exists
a limited range of the number of extra electrons added, inside which the superconductivity
can be stabilized if the structural and magnetic phase transitions are suppressed enough.
Moreover, comparison with pressure-temperature phase diagram data,
for samples spanning the whole doping range, further reenforces the conclusion that suppression
of the structural / magnetic phase transition temperature
enhances $T_c$ on the underdoped side, but for the overdoped side $T_C^{max}$
is determined by $e$.
Therefore,
by choosing the combination of dopants that are used, we can adjust
the relative positions of the upper phase lines (structural and magnetic phase transitions) and the
superconducting dome to control the occurrence and disappearance of the superconductivity in transition metal, electron-doped ${\rm BaFe}_2{\rm As}_2$.

\end{abstract}
\pacs{74.10.+v; 74.62.Dh; 74.70.Xa; 74.25.-q}
\date{\today}
\maketitle
\section{introduction}
The iron pnictide superconductors have been the focus of extensive
research since the layered $\rm{LaFeAsO_{0.9}F_{0.1}}$ was
reported superconducting around 26 K at ambient pressure \cite{jacs} and later at 43 K,
under applied pressures up to 4 GPa \cite{pressure1111}.
$T_c$ soon rose to 55 K
at ambient pressure in $\rm{RFeAsO_{0.9}F_{0.1}}$ (R=Ce, Pr, Nd,
Sm) \cite{ce1111, pr1111, nd1111, sm1111}.
But the size of single crystals of these 1111 superconductors, grown by either a high temperature / high pressure technique \cite{pressuregrowth}
or flux-growth method \cite{flux1111}, were small
and thus limited the research on the 1111 system. In addition, problems associated with the stoichiometry
of O and F made reproducibility hard to maintain in these
compounds.

Fortunately, another high $T_c$,
Fe-pnictide family with $T_c$ up to 38 K, $\rm {(Ba_{1-x}K_x)Fe_2As_2}$, was soon discovered \cite{rotter, rotter2}.
Following the discovery of this oxygen-free compound in polycrystalline form, sizable single crystals of $\rm{ (Ba_{1-x}K_x)Fe_2As_2}$
were grown, using solution growth methods, with dimensions up to $3\times 3\times 0.2$ $mm^3$ \cite{nik, wang1, chen2}.
Unfortunately these $K$-doped samples were found to be rather inhomogeneous and
there is a significant layer to layer concentration variation even in one piece
\cite{nik, kfeas}. On the other hand, it was soon
found that transition metal doping on the Fe site in this "122" family could induce superconductivity up to 24 K \cite{sefat,
nidope, rudoping, threedoping, nico}. This discovery was important not only because
it made Fe pnictides different from
cuprates in the sense that superconductivity is generally destroyed
by doping in the CuO plane, but also because
large, high quality, homogeneous single crystals could be easily grown and reproduced \cite{wang1, sefat, threedoping,  nico, chu, nmr, wangco, hall, ni4d}.
The crystal volume can be as large as 0.2 $cm^3$
and the samples are the most homogeneous ones among all the Fe pnictide superconductors,
which is critical for detailed and systematic studies. Given these advantages, even though $T_c$ is lower than 30 K,
a great deal of research has been done on these systems.

The phase diagram of ${\rm Ba(Fe}_{1-x}{\rm Co}_x)_2{\rm As}_2$ was mapped out in detail \cite{nico,chu, nmr, hall, wangco}.
It was found the structural and magnetic phase transitions are suppressed with doping and, at intermediate dopings,
superconductivity is stabilized with a coexistence range for
antiferromagnetism and superconductivity on the low $x$ side of the superconducting
dome \cite{threedoping, nico,chu, nmr, wangco, hall, ni4d, amesneutron,stanfordneutron, nandi, Fernandes}.

In order to compare the effects of $3d$ electron doping on ${\rm BaFe}_2{\rm As}_2$, and thus try to
discover the similarities and differences,
to understand the relation between the structural / antiferromagnetic phase transition and superconductivity, as well as the
conditions for the appearance of superconductivity in these systems,
we focus on electron doped ${\rm BaFe}_2{\rm As}_2$:
 ${\rm Ba(Fe}_{1-x}{\rm Ni}_x)_2{\rm As}_2$, ${\rm Ba(Fe}_{1-x}{\rm Cu}_x)_2{\rm As}_2$ ($x\leq0.356$) and two families of
${\rm Ba(Fe}_{1-x-y}{\rm Co}_x{\rm Cu}_y)_2{\rm As}_2$ ($x\sim0.022$ and $x\sim0.047$) series. Single crystals were grown
and characterized. An initial work which showed only the transport measurements on a subset of samples from
these series has been published \cite{threedoping}; in this paper a comprehensive study, on more samples and series, is presented.
In specific,
for these four series, data from structural, microscopic, transport and thermodynamic measurements are presented.
All these measurements show that the
structural / magnetic phase transitions at 134 $K$ in pure ${\rm BaFe}_2{\rm As}_2$ are monotonically suppressed and
separated by these dopants.
For the ${\rm Ba(Fe}_{1-x}{\rm Ni}_x)_2{\rm As}_2$ series, superconductivity is stabilized over
 a smaller doping range than that for the ${\rm Ba(Fe}_{1-x}{\rm Co}_x)_2{\rm As}_2$ series.
 High field, anisotropic $H_{c2}(T)$ measurements done on the optimally Ni doped ${\rm BaFe}_2{\rm As}_2$ sample,
with an applied magnetic field up to 33 T, revealed behavior comparable to that found for K- and
Co- doped ${\rm BaFe}_2{\rm As}_2$ \cite{nik,nico,mielke}.
In the ${\rm Ba(Fe}_{1-x}{\rm Cu}_x)_2{\rm As}_2$ ($x\leq0.356$) series,
superconductivity is not stabilized for $T>3$ K. For one concentration, $x=0.044$, a sharp drop of the resistivity to zero shows up near 2 $K$.
This feature may be a sign of \emph{very} limited ($0.035 < x < 0.050$) superconducting region near this Cu doping level.
In the ${\rm Ba(Fe}_{1-x-y}{\rm Co}_{x}{\rm Cu}_y)_2{\rm As}_2$ ($x\sim0.022$) series, although ${\rm Ba(Fe}_{0.976}{\rm Co}_{0.024})_2{\rm As}_2$ is
not superconducting, the introduction of extra Cu atoms further suppresses the structural / magnetic phase transitions and a
$T_c$ dome, with a maximum $T_c$ value of 12 K, is found.
In the ${\rm Ba(Fe}_{1-x-y}{\rm Co}_{x}{\rm Cu}_y)_2{\rm As}_2$ ($x\sim0.047$) series,
Cu is doped into ${\rm Ba(Fe}_{0.953}{\rm Co}_{0.047})_2{\rm As}_2$, an underdoped compound with $T_c\sim 16$ K.
As Cu is added, the structural and magnetic phase transitions
are suppressed further, and $T_c$ rises to $\sim$ 20 K.
Comparisons of the $T-x$ and $T-e$ phase diagrams for
TM=Co, Ni, Cu, Cu / Co series combined with our previous work on Rh, Pd dopings \cite{ni4d} reveal that, although the suppression of
the upper transitions better scales with the doping level $x$, the location and extent
of the superconducting dome scales better with the number of extra conduction
electrons added, which are one for each Co, two for each Ni and three for each Cu
atom.

\section{Sample growth, structural and compositional determination and experimental methods}

Single crystals of ${\rm Ba(Fe}_{1-x}{\rm TM}_x)_2{\rm As}_2$ (TM=Ni, Cu, Cu / Co) were grown out of a TMAs self flux, using
conventional high-temperature solution growth techniques \cite{canfield92}. The growth protocol
of Ni doped ${\rm BaFe}_2{\rm As}_2$ single crystal growths is the same as for Co-doping  \cite{nico}.
Cu doped ${\rm BaFe}_2{\rm As}_2$ and Cu / Co doped ${\rm BaFe}_2{\rm As}_2$ single crystal
growths are slightly different, though. We use small Cu shot, rather than
CuAs, to introduce the dopant because no binary CuAs compound is known to exist.
 For Cu doped ${\rm BaFe}_2{\rm As}_2$ ($x\leq 0.356)$, small Ba chunks, FeAs powder and Cu shot were mixed together
 according to the ratio Ba : FeAs : Cu = 1 : 4 : $m$. The nominal concentration $x_{nominal}$ can be calculated as Cu/(Cu+Fe) = $m/(4+m)$.
 For ${\rm Ba(Fe}_{1-x-y}{\rm Co}_{x}{\rm Cu}_y)_2{\rm As}_2$ ($x\sim 0.022$), small Ba chunks, FeAs, CoAs powder and Cu shot
were mixed together according to the ratio Ba : FeAs : CoAs : Cu = 1 : 3.88 : 0.12 : $m$.
For ${\rm Ba(Fe}_{1-x-y}{\rm Co}_{x}{\rm Cu}_y)_2{\rm As}_2$ ($x\sim0.047$), Ba : FeAs : CoAs : Cu = 1 : 3.75: 0.25 : $m$ were mixed.
These mixtures were placed into a 2 $ml$ or 5 $ml$ alumina crucible. A second, catch crucible,
containing quartz wool, was placed on top of this growth crucible and then both were
 sealed in a quartz tube under $\sim 1/3$ atmosphere Ar gas. The sealed quartz tube was heated up to 1180 $^\circ $C,
 stayed at 1180 $^\circ $C for 5 to 8 hours, and then cooled to 1000 $^\circ $C over 36 hours. Once the furnace reached
 1000 $^\circ $C, the excess liquid was decanted from the plate like single crystals.

Given the difficulties associated with K homogeneity \cite{nik, kfeas}, determining how homogeneous the TM doped samples are is important.
Using wavelength
dispersive x-ray spectroscopy (WDS) in the electron probe microanalyzer
of a JEOL JXA-8200 electron-microprobe, extensive elemental analysis was performed on each of these batches,
especially on the pieces which were used to make the
magnetization, resistivity and heat capacity measurements. For those pieces, the samples
were carefully exfoliated and cut into several pieces.
WDS measurements were done up to five pieces of sample from each batch.
The average $x$ and $y$ values,
\clearpage
\bet
\begin{tabular}{c c c c c c c c}
   \hline
   \hline

   $~$ & ~ & ~ & ~ & ~ & ~ & ~ \\
   \multicolumn{7}{c}{${\rm Ba(Fe}_{1-x}{\rm Ni}_x)_2{\rm As}_2$}\\
   $~$ & ~ & ~ & ~ & ~ & ~ & ~ \\
   \hline
   N & 18 & 10 & 44 & 18 & 11 & 12 & 28 \\

   $x_{nominal}$ & 0.01 & 0.02 & 0.03 & 0.04 & 0.05 & 0.07 & 0.09 \\

   $x_{WDS}$ & 0.0067 & 0.016 & 0.024 & 0.032 & 0.046 & 0.054 & 0.072 \\

   $2\sigma$ & 0.001 & 0.002 & 0.002 & 0.003 & 0.002 & 0.002 & 0.004 \\
   \hline
   \hline
   $~$ & ~ & ~ & ~ & ~ & ~ & ~ \\
   \multicolumn{7}{c}{${\rm Ba(Fe}_{1-x}{\rm Cu}_x)_2{\rm As}_2$}\\
   \hline

   N & 16 & 11 & 17 & 26 & 12 & 16 \\

   $m$ & 0.02 & 0.05 & 0.09 & 0.1 & 0.11 & 0.12 \\

   $x_{nominal}$ & 0.005~ & 0.012 & 0.022 & 0.024 & 0.027 & 0.029 \\

   $x_{WDS}$ & 0.0077~ & 0.02 & 0.026 & 0.035 & 0.044 & 0.05 \\

   $2\sigma$ & 0.002 & 0.002 & 0.002 & 0.004 & 0.002 & 0.002\\
   \hline

   N & 43 & 12 & 10 & 8 & 17 & 23\\

   $m$ & 0.14 & 0.16 & 0.26 & 0.45 & 1 & 3 \\

   $x_{nominal}$ & 0.034 & 0.038 & 0.061 & 0.101 & 0.20 & 0.429\\

   $x_{WDS}$ & 0.061  & 0.068 & 0.092 & 0.165 & 0.288 & 0.356 \\

   $2\sigma$ & 0.002 & 0.002 & 0.008 & 0.02 & 0.02 & 0.02 \\
   \hline
   \hline
   $~$ & ~ & ~ & ~ & ~ & ~ & ~ \\
   \multicolumn{7}{c}{${\rm Ba(Fe}_{1-x-y}{\rm Co}_{x}{\rm Cu}_y)_2{\rm As}_2~~~(x\sim 0.022$)}\\
    \hline
   N & 18 & 12 & 20 & 30 & 20 & 20 & 28 \\

   $x_{WDS}$ & 0.024 & 0.024 & 0.022 & 0.022 & 0.021 & 0.021 & 0.021  \\

   $2\sigma$ & 0.001 & 0.001 & 0.001 & 0.001 & 0.001 & 0.001 & 0.001 \\

   $m$ & 0 & 0.014 & 0.03 & 0.05 & 0.07 & 0.09 & 0.14\\

   $y_{nominal}$ & 0 & 0.0035 & 0.0074 & 0.012 & 0.017 & 0.022 & 0.034 \\

   $y_{WDS}$ & 0 & 0.005 & 0.01 & 0.019 & 0.026 & 0.032 & 0.043  \\

   $2\sigma$ &0  & 0.002 & 0.002 & 0.003 & 0.004 & 0.003 & 0.004 \\
   \hline

   \end{tabular}
\
\eet
\clearpage

\bet
\begin{tabular}{c c c c c c c c}
\multicolumn{7}{c}{Table I continued}\\
   \hline
   \hline

   $~$ & ~ & ~ & ~ & ~ & ~ & ~ \\
   \multicolumn{7}{c}{${\rm Ba(Fe}_{1-x-y}{\rm Co}_{x}{\rm Cu}_y)_2{\rm As}_2~~~(x\sim 0.047)$}\\
     \hline
   N & 7 & 8 & 37 & 36 & 7 & 41 \\

   $x_{WDS}$ & 0.047 & 0.051 & 0.047 & 0.047 & 0.045 & 0.045 \\

   $2\sigma$ & 0.002 & 0.002 & 0.003 & 0.002 & 0.002 & 0.002 \\

   $m$ & 0 & 0.001 & 0.05 & 0.09 & 0.12 & 0.15\\

   $y_{nominal}$ & 0 & 0.0025 & 0.012 & 0.022 & 0.029 & 0.036 \\

   $y_{WDS}$ & 0 & 0.0045 & 0.019 & 0.034 & 0.046 & 0.058 \\

   $2\sigma$ & 0 & 0.001 & 0.002  & 0.004 & 0.004 & 0.006 \\
   \hline
   \hline
   \end{tabular}
\caption{WDS data for all five series.
N is the number of locations measured in one batch,
$m$ is as described in the crystal growth method part, $y_{nominal}$ is calculated as $m/(4+m)$,
$x_{WDS}$ and $y_{WDS}$ are the average $x$ and $y$ values measured in one batch, $2\sigma$
is two times the standard deviation of the N values measured.}
\eet
measured
at several locations on the sample from WDS measurement, $x_{WDS}$ and $y_{WDS}$, are used in this paper
rather than $x_{nominal}$ and $y_{nominal}$.

Table I summarizes the results of the WDS measurements of the
${\rm Ba(Fe}_{1-x}{\rm TM}_x)_2{\rm As}_2$ (TM=Ni, Cu, Cu / Co) series.
N is the total number of spots measured for a given batch.
$x_{nominal}$ and $y_{nominal}$ are the nominal doping
concentrations. $x_{WDS}$ and $y_{WDS}$ are the average values of the N measurements for a given batch.
$m$ is the quantity of elemental Cu added, as described above.
 $2\sigma$
is twice the standard deviation of the N values measured for one batch, which is taken as the compositional error bar in this paper.
The $2\sigma$ error bars, which also include machine errors, for all the spots measured in one batch are
$\lesssim$ 10\% of the average $x$ values. These results further demonstrate the relative
homogeneity of the ${\rm Ba(Fe}_{1-x}{\rm TM}_x)_2{\rm As}_2$ series.

\bef
\psfig{file=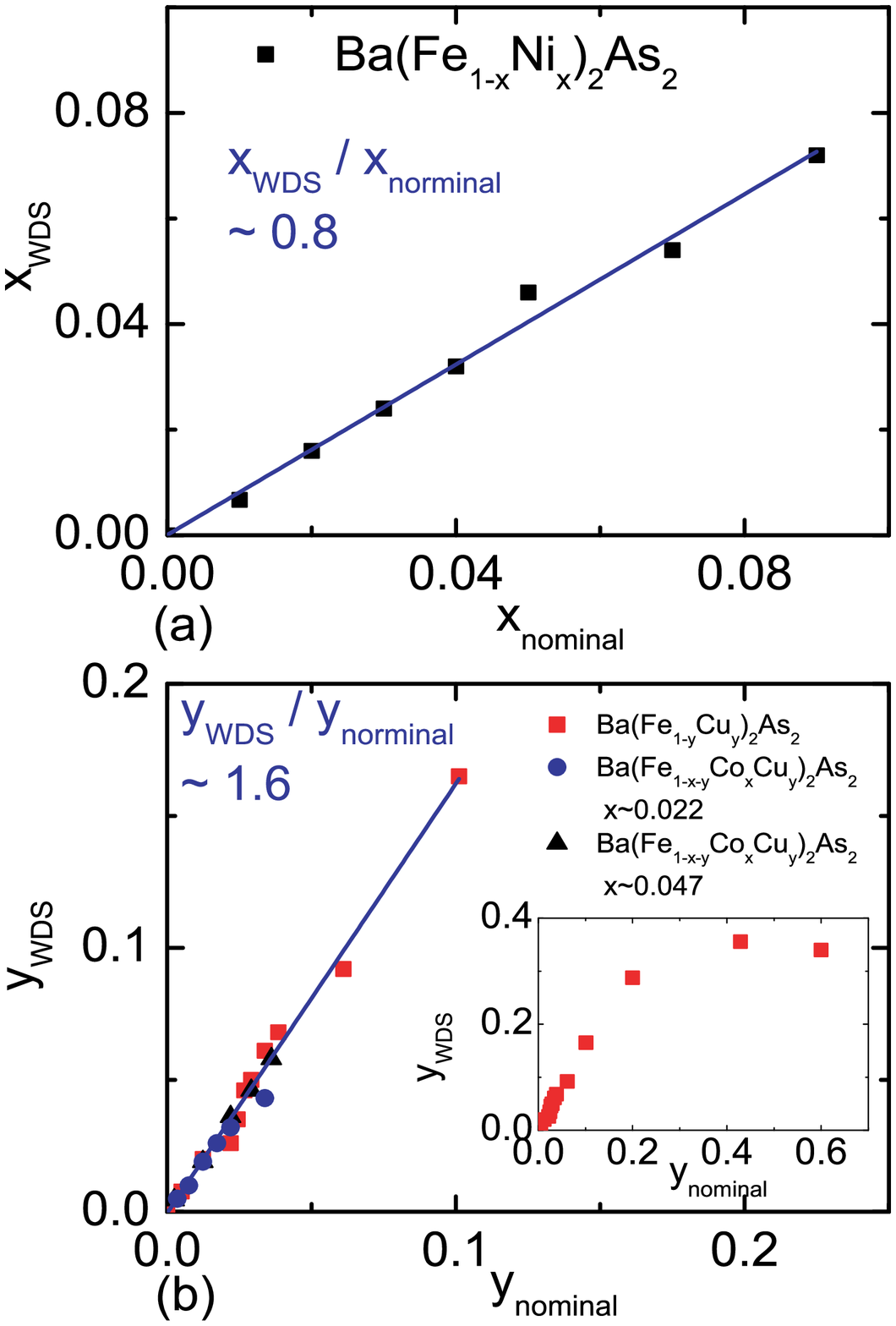,width=3.5in}
\caption{
(a) Measured Ni concentration vs. nominal Ni concentration for the ${\rm Ba(Fe}_{1-x}{\rm Ni}_x)_2{\rm As}_2$ series.
(b) Enlarged measured Cu concentration vs. nominal Cu concentration for
${\rm Ba(Fe}_{1-x-y}{\rm Co}_x{\rm Cu}_y)_2{\rm As}_2$ ($x=0$, $x \sim 0.022$ and $x \sim 0.047$). Inset:
$x_{WDS}$ vs. $x_{nominal}$ for ${\rm Ba(Fe}_{1-x}{\rm Cu}_x)_2{\rm As}_2$ in the whole doping range.}
\label{yield}
\eef

Figure \ref{yield} shows a graphic summary of the measured doping concentration vs. nominal doping concentration.
The data points for ${\rm Ba(Fe}_{1-x}{\rm Ni}_x)_2{\rm As}_2$
can be fitted well by a straight line. The ratio of the measured Ni concentration to the nominal Ni concentration is roughly 0.8.
For comparison, this number is 0.74 for Co doped ${\rm BaFe}_2{\rm As}_2$ \cite{nico}.
Figure \ref{yield} (b) summarizes the measured Cu concentration vs. nominal Cu concentration for low Cu dopings ($y_{norminal} < 0.1$) for
all ${\rm Ba(Fe}_{1-x}{\rm Cu}_x)_2{\rm As}_2$ and
${\rm Ba(Fe}_{1-x-y}{\rm Co}_x{\rm Cu}_y)_2{\rm As}_2$ ($x \sim 0.022$ and $x \sim 0.047$) growths.
Although Cu was doped in different series, all the data points fall on the same line.
The measured Cu concentration is roughly 1.6 times the nominal Cu concentration in this low doping range. For larger
Cu doping values, the ratio of WDS measured Cu concentration
 over nominal Cu concentration
decreases and the $x_{WDS}$ value saturates around 0.35, as shown in the inset of Fig. \ref{yield} (b).
 This could be due to the increasing TM:As ratio as $x_{Cu}^{nominal}$ increases; due to the use of Cu rather than CuAs, the TM : As
 ratio in the melt is 1.4 : 1 when
$x_{Cu}^{nominal}$=0.4 and 1.6 : 1 when $x_{Cu}^{nominal}$=0.6, both of which are much larger than
the value of 1:1 used for TM = Co and Ni.

Powder X-ray diffraction measurements, with a Si standard, were performed at room temperature on a Rigaku Miniflex diffractometer
with Cu $K_{\alpha}$ radiation.
Diffraction patterns were taken on ground single crystals from each batch.
The unit cell parameters were refined by "UnitCell"
software.
Peak positions were determined from the peak maximum.
Zero shift was corrected by the average shift of those Si peaks which have no overlap with the sample peaks.
Error bars were taken as twice of the standard deviation, $\sigma$,
which was obtained from the refinements.

\bef
\psfig{file=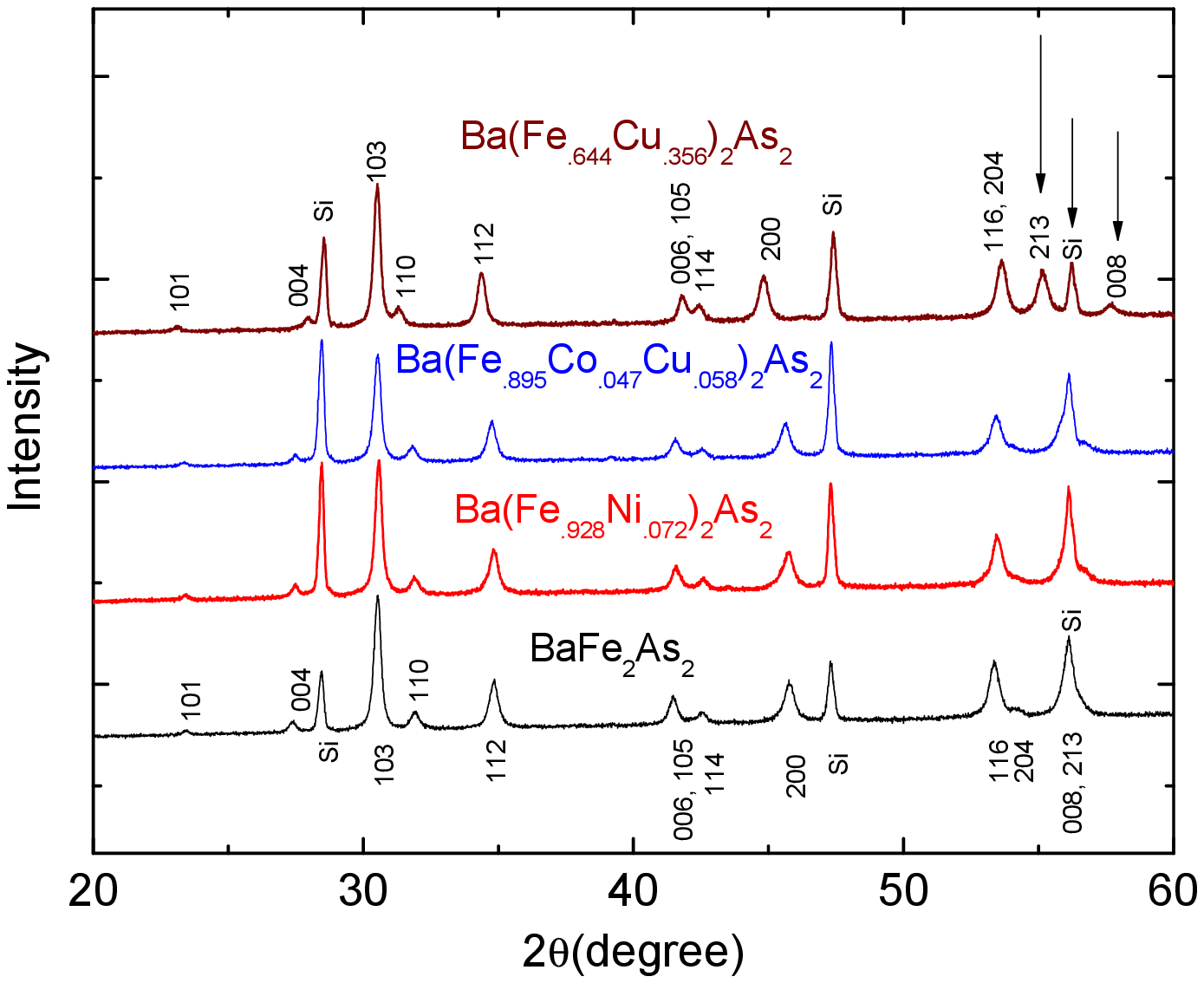,width=4in}
\caption{Powder X-ray diffraction patterns for ${\rm Ba(Fe}_{0.644}{\rm Cu}_{0.356})_2{\rm As}_2$, ${\rm Ba(Fe}_{0.895}{\rm Co}_{0.047}{\rm Cu}_{0.058})_2{\rm As}_2$
, ${\rm Ba(Fe}_{0.928}{\rm Ni}_{0.072})_2{\rm As}_2$ and pure ${\rm BaFe}_2{\rm As}_2$.}
\label{lattice}
\eef

Figure \ref{lattice} shows the powder x-ray diffraction patterns for pure ${\rm BaFe}_2{\rm As}_2$ and the samples which have
the highest doping level for each dopant:
${\rm Ba(Fe}_{0.644}{\rm Cu}_{0.356})_2{\rm As}_2$, ${\rm Ba(Fe}_{0.985}{\rm Co}_{0.047}{\rm Cu}_{0.058})_2{\rm As}_2$ and
${\rm Ba(Fe}_{0.928}{\rm Ni}_{0.072})_2{\rm As}_2$.
No impurity phases can be detected in any of these batches. Since ${\rm Ba(Fe}_{0.644}{\rm Cu}_{0.356})_2{\rm As}_2$ has the highest
doping concentration among all series, the lattice parameters manifest the largest changes;
the combined (213) and (008) peaks in pure ${\rm BaFe}_2{\rm As}_2$ that overlap the
 Si peak around 56${^\circ}$
split to either side and reveal three peaks which are indicated by arrows
in ${\rm Ba(Fe}_{0.644}{\rm Cu}_{0.356})_2{\rm As}_2$.

Heat capacity data were collected in a Quantum Design (QD) Physical Properties Measurement System (PPMS)
using the relaxation technique.
Magnetization and temperature-dependent AC electrical transport data (f=16 Hz, I=3 mA) were collected in a
QD Magnetic Properties Measurement System (MPMS) using a LR700 AC resistance bridge. Electrical contact was made to the sample by using Epotek
H20E silver epoxy to attach Pt wires in a four-probe configuration.
For all series, the measured room temperature resistivities varied from
$0.1 m\Omega~cm$ to $1 m\Omega~cm$. Because these samples are easy
to exfoliate or crack \cite{nico, tanatar1, tanatar2}, $\rho(T)/\rho_{300K}$
instead of resistivity is plotted as a function of temperature for all series in this paper.

Field-dependent DC electrical transport data were collected in the 33 T magnet facility
in National High Magnetic
Field Lab (NHMFL) in Tallahassee, FL.
$R~(H)$ data at different temperatures were measured for $H || c$ axis and $H \bot c$ axis.
To correct the temperature off-sets associated with the resistive probe used at the NHMFL \cite{nico}, $R(T)$ data for
both samples, in zero field, were measured in the Quantum Design MPMS unit. These shifts were at most 4\% of $T_c$.

\section{results}

\subsection{${\rm Ba(Fe}_{1-x}{\rm Ni}_x)_2{\rm As}_2$}

\bef \psfig{file=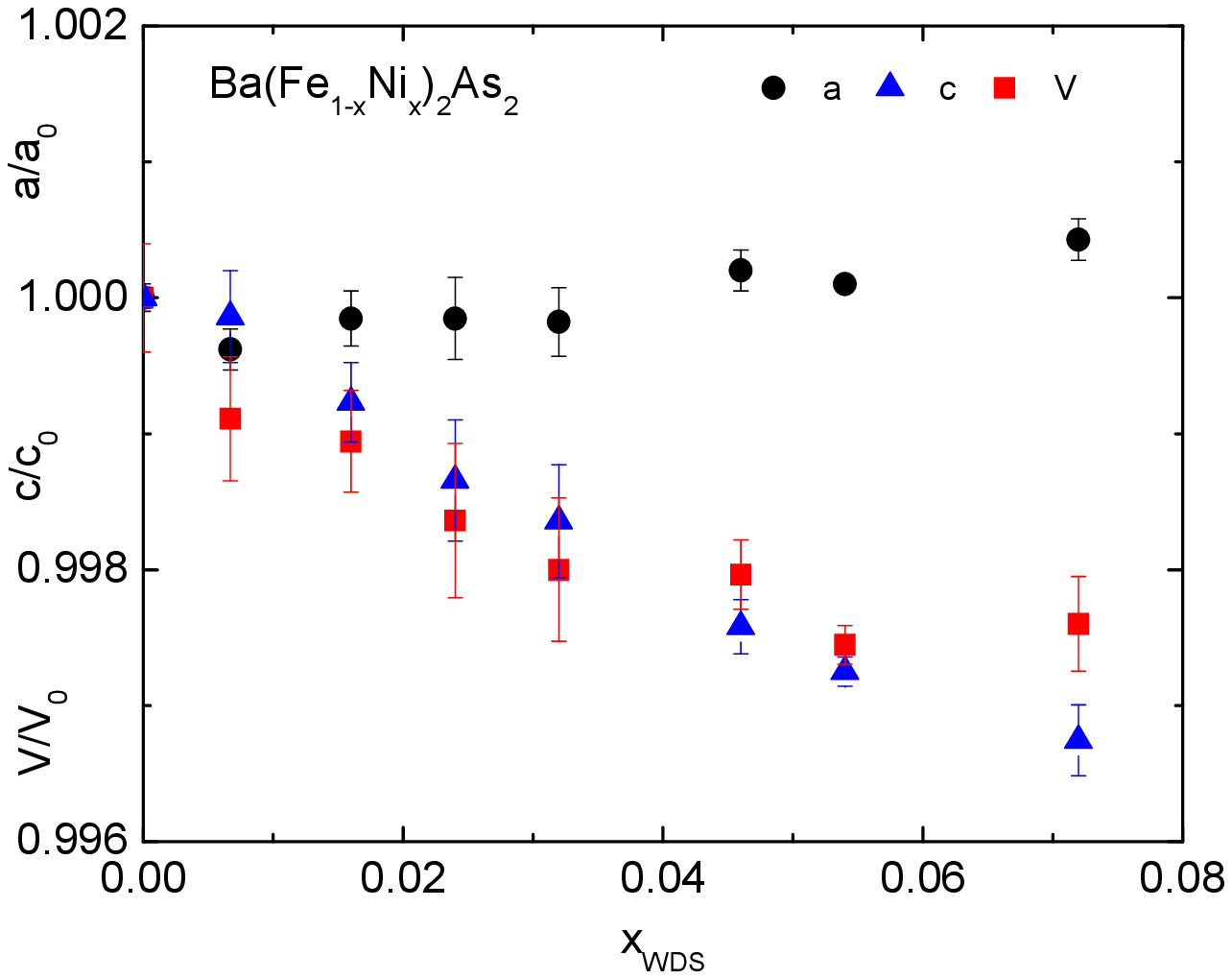,width=4in} \caption
{Room temperature lattice parameters of the ${\rm Ba(Fe}_{1-x}{\rm Ni}_x)_2{\rm As}_2$ series,
$a$ and $c$ as well as unit cell volume, $V$, normalized to
 the values of pure BaFe$_2$As$_2$ ($a_0$ = 3.9621(8) $\AA$, $c_0$ = 13.018(2) $\AA$,
$V_0$ = 204.357 $\AA^3$) as
 a function of measured Ni concentration, $x_{WDS}$
} \label{lni} \eef

{${\rm Ba(Fe}_{1-x}{\rm Ni}_x)_2{\rm As}_2$} compounds were reported to superconduct by Xu. et al \cite{nidope},
however, no detailed presentation of transport and thermodynamic data or determination of a phase diagram of the
structural, magnetic and superconducting phases was made. In order to
map the phase diagram of ${\rm Ba(Fe}_{1-x}{\rm Ni}_x)_2{\rm As}_2$, single crystals were grown and characterized.

The evolution of the lattice parameters with the doping level is shown in Fig. \ref{lni}.
For Ni dopings up to $x=$0.072, the
lattice parameter $a$ increases slightly, by 0.04\%, while the lattice parameter $c$ decreases almost ten times
faster, by 0.35\%, and thus the unit cell
volume decreases monotonically by 0.26\%.
This is different from Co-doped ${\rm BaFe}_2{\rm As}_2$, in which, up to the $x=0.114$ doping level,
$a$ and $c$ lattice parameters decrease by 0.07\%
and 0.5\% respectively and the unit cell volume decreases by 0.6\%.

\bef
\psfig{file=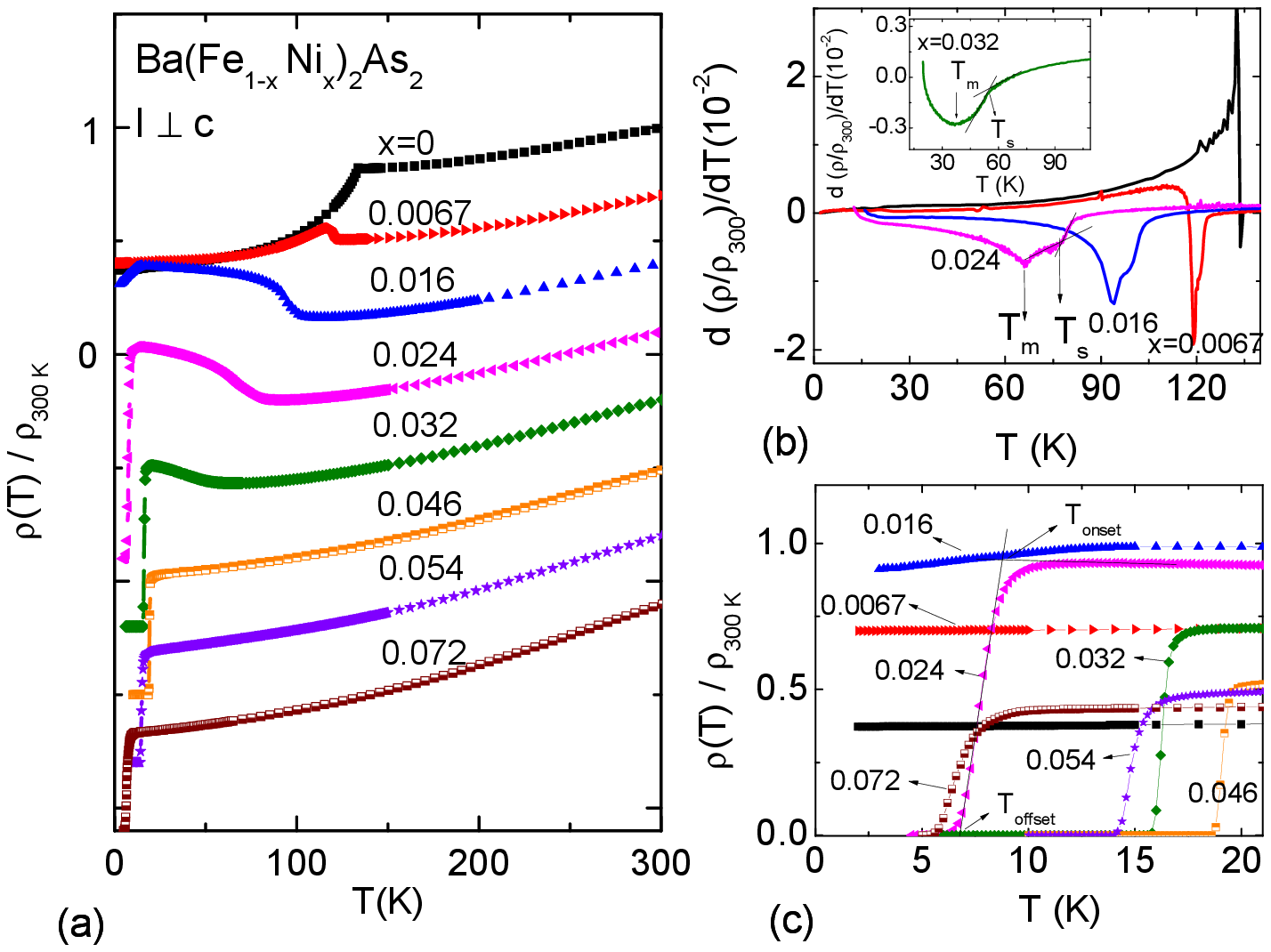,width=4in}
\caption{The ${\rm Ba(Fe}_{1-x}{\rm Ni}_x)_2{\rm As}_2$ series:
(a) The temperature
dependent resistivity, normalized to the room temperature value. Each subsequent data set is shifted downward by 0.3 for clarity.
(b) $d(\rho(T)/\rho_{300K})/dT$ for $y \leq 0.032$. The criteria to infer $T_s$ and $T_m$ from the resistivity data are shown.
(c) Enlarged low temperature $\rho(T)/\rho_{300K}$. The offset and onset criteria to infer $T_c$ are shown.
}
\label{rni}
\eef

Figure \ref{rni} (a) shows the normalized resistivity data taken from 2 K to 300 K for ${\rm Ba(Fe}_{1-x}{\rm Ni}_x)_2{\rm As}_2$.
Each subsequent data set is shifted downward by 0.3 for clarity.
The data show behavior very similar to ${\rm Ba(Fe}_{1-x}{\rm Co}_x)_2{\rm As}_2$ \cite{nico, chu, hall}. With Ni doping, the resistive anomaly associated with
the structural and magnetic phase transitions is suppressed
from 134 K, for pure ${\rm BaFe}_2{\rm As}_2$, to lower temperatures. For the lowest doping level, $x=0.0067$,
the resistive anomaly is very
similar to that seen in pure ${\rm CaFe}_2{\rm As}_2$ \cite{nica}
as well as very lightly Co doped ${\rm BaFe}_2{\rm As}_2$ \cite{nico}.
With higher Ni doping, the resistive anomaly becomes a broadened upturn. The suppression of the resistive anomaly can also be seen in Fig. \ref{rni} (b),
which shows the
enlarged $d(\rho(T)/\rho_{300K})/dT$ below 140 K for ${\rm Ba(Fe}_{1-x}{\rm Ni}_x)_2{\rm As}_2$;
two kinks similar to those in ${\rm Ba(Fe}_{1-x}{\rm Co}_x)_2{\rm As}_2$ \cite{nico, amesneutron} can be observed.
Based on the Co-doping work \cite{nico, amesneutron}, and considering the similarities between Co-doping and Ni-doping, it is natural to believe that
 the higher-temperature feature is associated with
 the structural phase transition and the lower-temperature feature is associated with the magnetic phase transition.
Recent neutron scattering work \cite{andreas}
on the ${\rm Ba(Fe}_{0.961}{\rm Rh}_{0.039})_2{\rm As}_2$ compound has confirmed this assumption and clarified the criteria to
infer the structural phase transition temperature ($T_s$) and magnetic phase transition temperature ($T_m$) from the resistivity data,
which are shown for $x=0.024$ sample in
Fig. \ref{rni} (b). These criteria will be employed
in this paper for the samples which have two distinct kinks in $d(\rho(T)/\rho_{300K})/dT$ (including Co-doping).
For the samples which have blurred kinks in $d(\rho(T)/\rho_{300K})/dT$
due to the nearness between $T_s$, $T_m$ and the superconducting temperature, $T_c$, like the $x=0.032$ sample, the criteria to infer $T_s$ and $T_m$ are shown
in the inset
of Fig. \ref{rni} (b).

As we can see, as $T_s$ and $T_m$ are suppressed, superconductivity appears.
For $x=0.024$, $T_s$ is suppressed to 77 K, $T_m$ is suppressed to 66 K,
and zero resistivity is detected below 6.8 K. For $x=0.046$, the resistive anomaly associated with structural and magnetic phase transitions
is no longer detected and $T_c$ increases to the maximum value around 19 K. For larger $x$, $T_c$ decreases
and is suppressed to $\sim$ 5.7 K for $x=0.072$. The superconducting feature can be seen more clearly in Fig. \ref{rni} (c),
which presents the low temperature resistivity
data for ${\rm Ba(Fe}_{1-x}{\rm Ni}_x)_2{\rm As}_2$. The offset and onset criteria to determine $T_c$,
are also shown in Fig. \ref{rni} (c). These criteria are employed to infer $T_c$ from resistivity data in this paper.
It can be seen that the superconducting transition width of ${\rm Ba(Fe}_{1-x}{\rm Ni}_x)_2{\rm As}_2$
is smaller than 2 K as inferred from the resistivity measurements.

\bef
\psfig{file=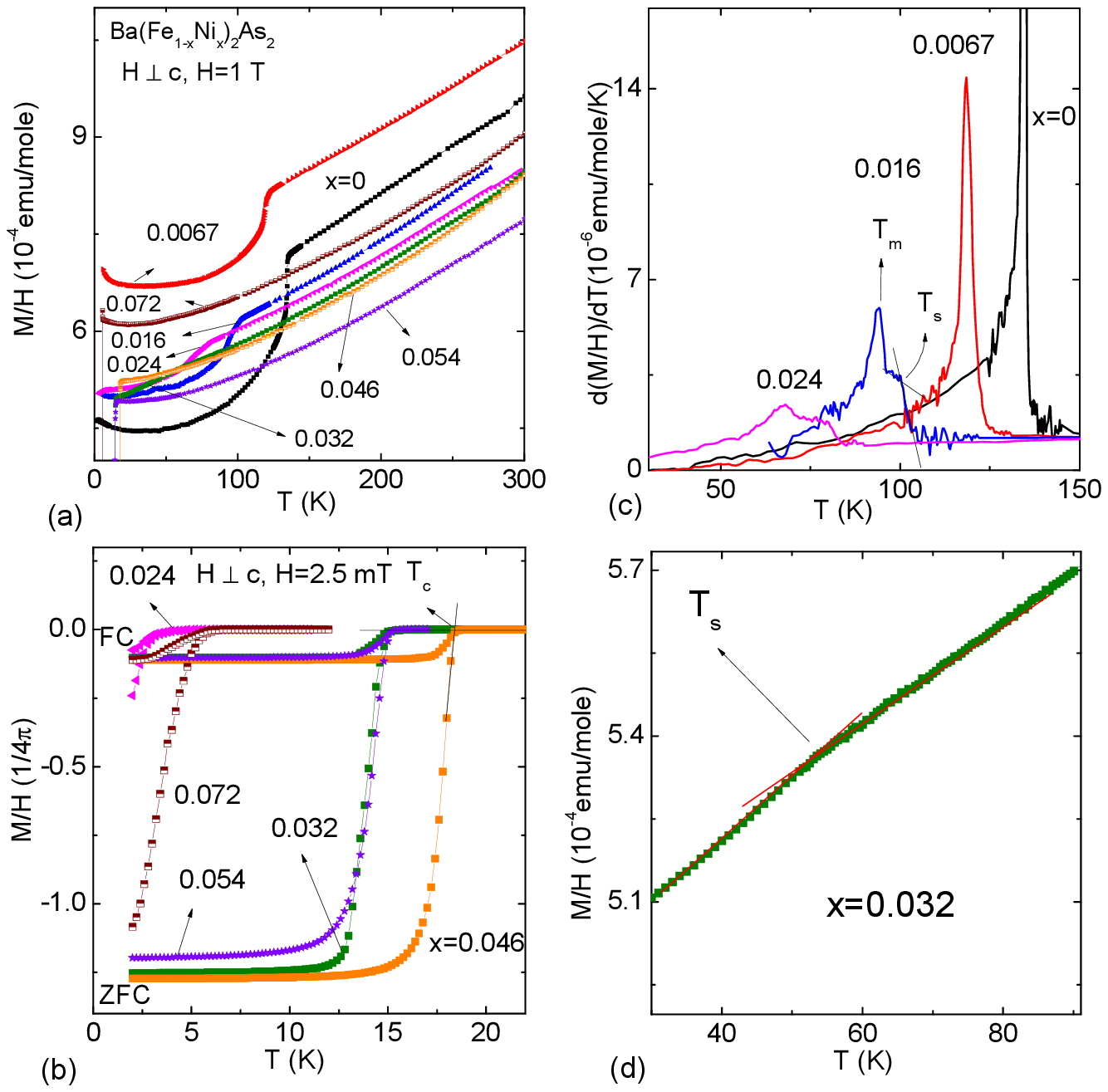,width=5in}
\caption{The ${\rm Ba(Fe}_{1-x}{\rm Ni}_x)_2{\rm As}_2$ series:
(a) $M(T)/H$ data taken at 1 T with $H\bot c$.
(b) Field-cooled (FC) and zero-field-cooled (ZFC) low field $M(T)/H$ data taken at 2.5 mT with $H\bot c$.
The criterion to infer $T_c$ is shown.
(c) $d(M(T)/H)/dT$ for $x \leq 0.024$. The criteria to infer $T_s$ and $T_m$ are shown.
(d) The criterion to infer $T_s$ for $x=0.032$ sample.
}
\label{mni}
\eef

Figure \ref{mni} (a) shows the ${M(T)/H}$
data taken in 1 T with $H\bot c$. For pure ${\rm BaFe}_2{\rm As}_2$, a drop in
susceptibility, associated with the structural / magnetic phase transitions
around 134 K, can be clearly seen. With Ni doping, this feature is suppressed to lower temperature and the derivative, $d(M/H)/dT$,
 presented in Fig. \ref{mni} (c), splits,
consistent with the resistivity data shown in Fig. \ref{rni} (b). The criteria to infer $T_s$ and $T_m$ from the magnetization data are shown
in Fig. \ref{mni} (c) and were employed in this paper for the samples which have two distinct kinks in $d(M/H)/dT$.
For sample $x=0.032$, due to the nearness of $T_s$, $T_m$ and $T_c$, only a very weak slope change can be detected in the magnetization data, therefore
the criterion to infer $T_s$ is different and shown in Fig. \ref{mni} (d). There is no detectable feature for us to infer $T_m$ from
the magnetization data for $x=0.032$ sample.
It is also worth noting that $M(T)/H$ data show an almost linear temperature dependence
above the structural and magnetic phase transition temperatures. This linear behavior is
similar to that seen in ${\rm Ba(Fe}_{1-x}{\rm Co}_x)_2{\rm As}_2$ \cite{nico, wangco}.
The magnitude of the susceptibility in the normal state, centered around $7\times 10^{-4}$ emu/mole,
is again similar to that of ${\rm Ba(Fe}_{1-x}{\rm Co}_x)_2{\rm As}_2$.
Figure \ref{mni} (b) shows the ${M(T)/H}$ data taken at 2.5 mT with H perpendicular to the
crystallographic $c$-axis of the ${\rm Ba(Fe}_{1-x}{\rm Ni}_x)_2{\rm As}_2$ samples.
The Meissner effect can be clearly seen in the field-cooled (FC) data, the zero-field-cooled (ZFC) data highlight the transition even more dramatically.
The supercondcuting fractions
are similar to the superconducting fractions of ${\rm Ba(Fe}_{1-x}{\rm Co}_x)_2{\rm As}_2$ \cite{nico}.
The criterion to determine $T_c$ from the magnetization data is shown
for $x=0.046$ sample which has the maximum $T_c$ in this series, and
will be used for all the series presented in this paper.

The heat capacity data of the ${\rm Ba(Fe}_{1-x}{\rm Ni}_x)_2{\rm As}_2$ series have been presented and published in reference \cite{sergey}.
Together with the heat capacity data of the ${\rm Ba(Fe}_{1-x}{\rm Co}_x)_2{\rm As}_2$ series, a $\Delta C/T_c \propto T_c^2$ relation was revealed.

The structural / magnetic and superconducting transition temperatures inferred
from Figs. \ref{rni}, \ref{mni} and the heat capacity data \cite{sergey} are summarized in Table \ref{tableni} and Fig. \ref{phaseni}.
The criteria to infer these temperature are shown in Fig. \ref{rni} (b) and Fig. \ref{mni} (b).
For $x=0.032$ sample, $T_s$ and $T_m$ are marked with * in the table since different criteria are employed for this concentration.
As we can see from Table. \ref{tableni}, for small $x$ values, $T_s$ and $T_m$ are suppressed and split.
\bet
\begin{tabular}{c | c | c c c c | c c c | c c c}
   \hline
   \hline

    dopant & $x$ & \multicolumn{4}{|c|}{$\rho$}& \multicolumn{3}{|c|}{$M$}& \multicolumn{1}{c}{$C$}&\\
    \hline
   Ni&~ & {$T_s$}& {$T_m$}& {$T_c^{onset}$}& {$T_c^{offset}$} &  {$T_s$}& {$T_m$}& {$T_c$} & {$T_c$}\\
 ~                  &0 & 134 & 134 &     &     & 134    &134 &    &  \\
  ~                  &0.0067 & 121 & 118 &     &     & 119    & 119 &    &  \\

  ~                  &0.016 & 100   & 94   &     &     &100  &94   &    &  \\

   ~                  &0.024 & 77   & 66  &8.6  &6.8  &80 & 68  &3.9 & 2.5   \\

    ~                &0.032 & 54*     &$37^{*}$       &16.6 &15.9 & 53*  &       &15.1 &14.6   \\

     ~                &0.046 &        &       &19.4 &18.8 &     &       &18.4 &17.8    \\

      ~                &0.054 &        &       &15.5 &14.3 &     &       &14.4 &13.9    \\

      ~                  &0.072 &        &       &7.5  &5.7   &    &       &6    &5.2       \\
\hline
   \hline
  \end{tabular}

\caption{Summary of $T_s$, $T_m$ and $T_c$ from resistivity, magnetization and specific heat measurements for the
 ${\rm Ba(Fe}_{1-x}{\rm Ni}_x)_2{\rm As}_2$ series. *: see text.}
 \label{tableni}
\eet
\bef
\psfig{file=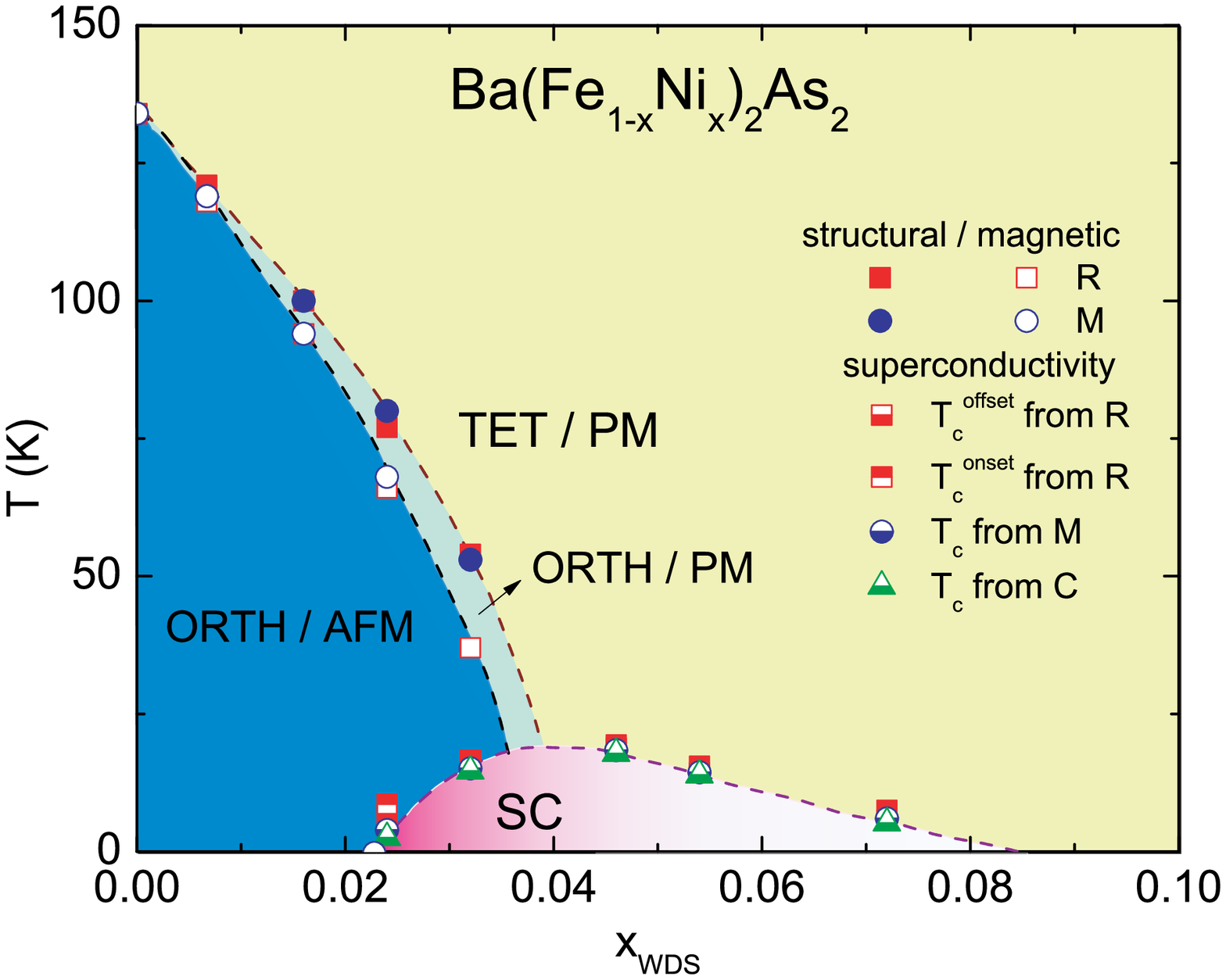,width=4in}
\caption{$T - x$ phase diagram of
Ba(Fe$_{1-x}$Ni$_x$)$_2$As$_2$ single crystals for $x \leq 0.072$.
The precise form of $T_s$ and $T_m$ lines are not yet determined in the superconducting
dome region, but we assume that they intersect with the superconducting dome near
$T_c^{max}$ \cite{nandi}, which is implied by the shading plotted in the superconducting dome.
}\label{phaseni}
\eef
For higher $x$-values, superconductivity is stabilized as
$T_s$ and $T_m$ continue to be suppressed.
 All of the $T-x$ data can be used to assemble a temperature-doping concentration ($T-x$) phase diagram for
${\rm Ba(Fe}_{1-x}{\rm Ni}_x)_2{\rm As}_2$ as shown in Fig. \ref{phaseni}. It has very similar appearance as
the one for ${\rm Ba(Fe}_{1-x}{\rm Co}_x)_2{\rm As}_2$ except the superconducting dome occurs at a lower $x$ and over a smaller $x$-range.

\bef
\psfig{file=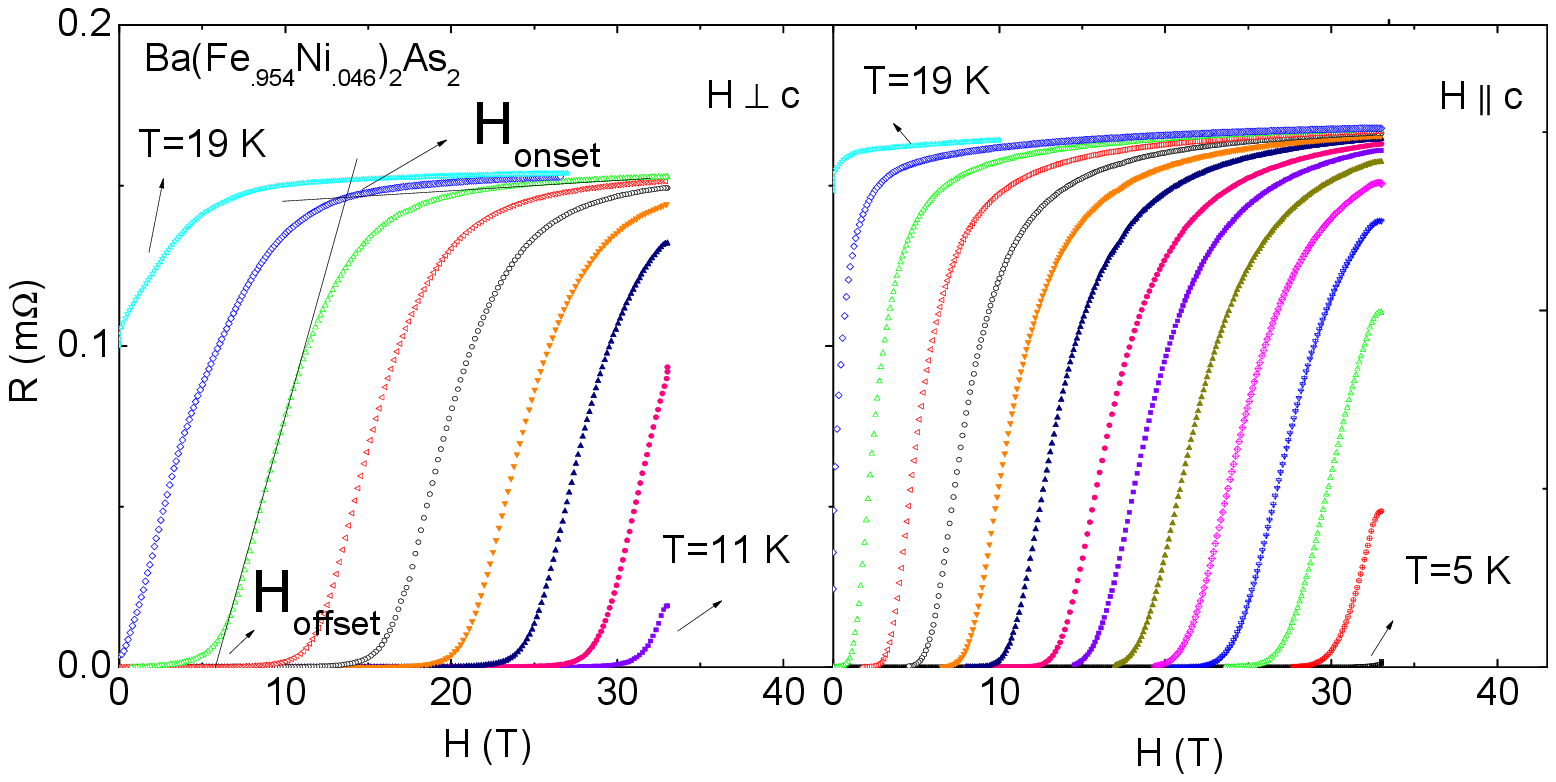,width=5in}
\caption{$R(H)$ data of
${\rm Ba(Fe}_{0.954}{\rm Ni}_{0.046})_2{\rm As}_2$ with $H\bot c$ (left panel) and $H || c$ (right panel).
}
\label{hc2ni}
\eef

Given the similarities, and differences,
between the Ni doped and Co doped ${\rm BaFe}_2{\rm As}_2$ systems, a comparison of the $H_{c2}(T)$ curves, which
reflect the properties of the superconductivity in these two systems, is desirable.
Anisotropic $H_{c2}$ data taken for ${\rm Ba(Fe}_{0.954}{\rm Ni}_{0.046})_2{\rm As}_2$
in the magnetic fields up to 33 T are summarized in Fig. \ref{hc2ni}. Although data
was taken on two samples, only one set of R(H) data is shown.
The left panel of Fig. \ref{hc2ni} presents the $R(H)$ data taken from 11 K to 19 K in 1 K steps for $H\bot c$.
The right panel presents the $R(H)$ data taken from 5 K to 19 K in 1 K steps for $H || c$.
Offset and onset criteria to infer $H_{c2}$ are shown.

Two Co dopings are logically comparable to the near optimally
doped ${\rm Ba(Fe}_{0.954}{\rm Ni}_{0.046})_2{\rm As}_2$: the comparably doped ${\rm Ba(Fe}_{0.953}{\rm Co}_{0.047})_2{\rm As}_2$
which has a similar $T_c$, and the near-optimally doped ${\rm Ba(Fe}_{0.926}{\rm Co}_{0.074})_2{\rm As}_2$.
Temperature dependent $H_{c2}$ curves for ${\rm Ba(Fe}_{0.954}{\rm Ni}_{0.046})_2{\rm As}_2$
 are presented in Fig. \ref{hc2ni1} in comparison with
${\rm Ba(Fe}_{0.953}{\rm Co}_{0.047})_2{\rm As}_2$ (Fig. \ref{hc2ni1} (a)) and ${\rm Ba(Fe}_{0.926}{\rm Co}_{0.074})_2{\rm As}_2$
(Fig. \ref{hc2ni1} (b)).
The anisotropy of near-optimally doped ${\rm Ba(Fe}_{0.954}{\rm Ni}_{0.046})_2{\rm As}_2$ is virtually identical to near-optimally doped
${\rm Ba(Fe}_{0.926}{\rm Co}_{0.074})_2{\rm As}_2$ as indicated from Fig. \ref{hc2ni1} (b) whereas it is
almost 2 times larger than that of the underdoped ${\rm Ba(Fe}_{0.953}{\rm Co}_{0.047})_2{\rm As}_2$ (similar
doping level, similar $T_c$) as shown in Fig. \ref{hc2ni1} (a).
This is a clear manifestation of the idea that the anisotropy of the superconducting state is not defined by $x$, but rather by the low temperature
structural / magnetic state of the system \cite{nico}. The anisotropic parameter
$\gamma$ (= $H_{c2}^{\bot c}(T)/H_{c2}^{||c}(T)$) of ${\rm Ba(Fe}_{0.954}{\rm Ni}_{0.046})_2{\rm As}_2$
is shown in Fig. \ref{hc2ni1} (c). It was calculated by taking each data point from
$H_{c2}^{\bot c}(T)$ curve and interpolating $H_{c2}^{||c}(T)$ at the same T-value, from the $H_{c2}^{||c}$ curve.
As we can see, $\gamma$ varies from 2 far from $T_c$ to 3 near to $T_c$ by offset criterion or
from 1.7 far from $T_c$ to 3 near to $T_c$ by onset criterion.

\bef
\psfig{file=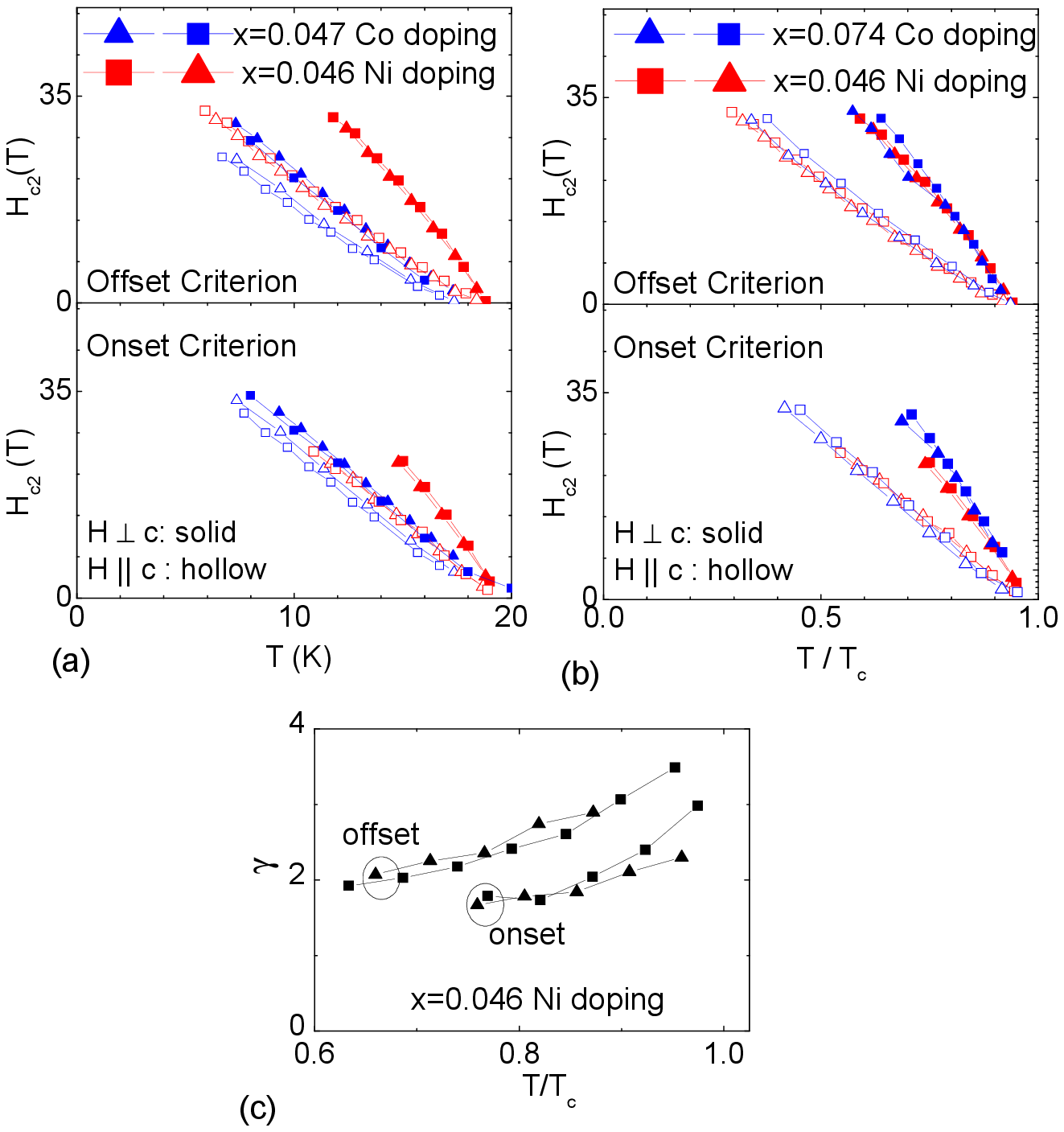,width=5in}
\caption{(a) $H_{c2}$ vs. $T$ from offset criterion (upper panel) and onset criterion (lower panel) of ${\rm Ba(Fe}_{0.954}{\rm Ni}_{0.046})_2{\rm As}_2$
and ${\rm Ba(Fe}_{0.953}{\rm Co}_{0.047})_2{\rm As}_2$ \cite{nico}.
(b) $H_{c2}$ vs. $T/T_c$ from offset criterion (upper panel) and onset criterion (lower panel) of ${\rm Ba(Fe}_{0.954}{\rm Ni}_{0.046})_2{\rm As}_2$
and ${\rm Ba(Fe}_{0.926}{\rm Co}_{0.074})_2{\rm As}_2$ \cite{nico}.
(c) $\gamma=H_{c2}^{\bot c}/H_{c2}^{|| c}$ vs. $T/T_c$ for ${\rm Ba(Fe}_{0.954}{\rm Ni}_{0.046})_2{\rm As}_2$. For each composition, data
inferred from $R(H)$ measurements on two samples are shown.}
\label{hc2ni1}
\eef

Considering two samples and two criteria, for ${\rm Ba(Fe}_{0.954}{\rm Ni}_{0.046})_2{\rm As}_2$,
 $(dH_{c2}^{||c}/dT)|_{T_c}$ ranges from -2.2 to -3 $T/K$ and
$(dH_{c2}^{\bot c}/dT)|_{T_c}$ ranges from -5 to -5.7 $T/K$.
Assuming the validity of Werthamer-Helfand-Hohenberg (WHH) equation, $H_{c2}(0)= -0.693T_c(dH_{c2}/dT)|_{T_c}$, $H_{c2}^{\bot c}(0)$ can be estimated
to be 70 T to 80 T and $H_{c2}^{||c}(0)$ can be between 30 T to 40 T. Using the equations
$\xi^{\bot c} = (\phi_0/2\pi H_{c2}^{||c})^{1/2}$ and
$\xi^{||c} = (\phi_0 H_{c2}^{||c}/2\pi (H_{c2}^{\bot c})^2)^{1/2}$,
the coherence length of in plane
$\xi^{\bot c}(0)$ is around 30 $\AA$ and inter plane $\xi^{||c}(0)$ is around 14 $\AA$.
Alternatively, given that the anisotropic $H_{c2}(T)$ data for optimally Co and Ni doped ${\rm BaFe}_2{\rm As}_2$ is similar to that
found for K-doped ${\rm BaFe}_2{\rm As}_2$ \cite {nik}, it is likely that $H_{c2}^{\bot c}(T)$ will continue to bend over to meet the essentially
linear $H_{c2}^{||c}(T)$ curve near $H_{c2}(0)$ $\sim$ 50 T, giving an isotropic coherence length of 26 $\AA$.

\subsection{ ${\rm Ba(Fe}_{1-x}{\rm Cu}_x)_2{\rm As}_2$}

\bef \psfig{file=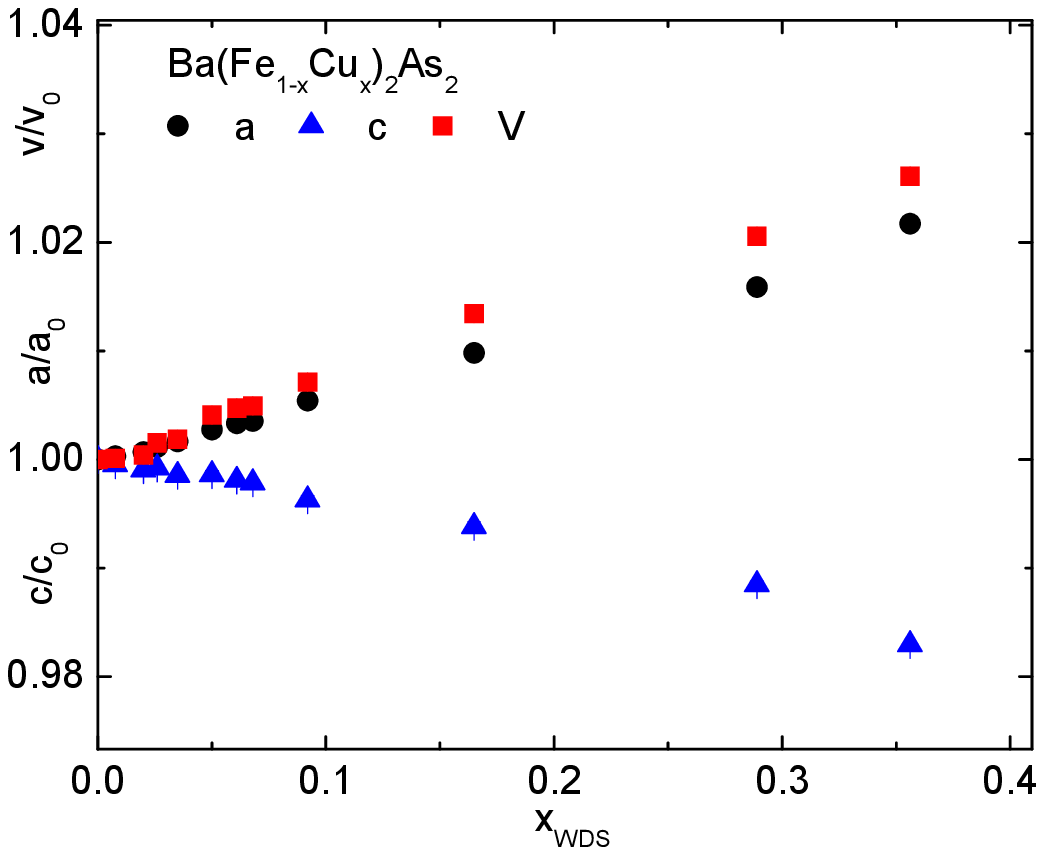,width=4in} \caption{
Lattice parameters of the ${\rm Ba(Fe}_{1-x}{\rm Cu}_x)_2{\rm As}_2$ series, $a$ and $c$ as well as unit cell volume, $V$, normalized to
 the values of pure BaFe$_2$As$_2$ ($a_0$ = 3.9621(8) $\AA$, $c_0$ = 13.0178 (20) $\AA$,
$V_0$ = 204.357 $\AA^3$) as
 a function of measured Cu concentration, $x_{WDS}$
} \label{lcu} \eef
Since superconductivity was found in both ${\rm Ba(Fe}_{1-x}{\rm Co}_x)_2{\rm As}_2$ and ${\rm Ba(Fe}_{1-x}{\rm Ni}_x)_2{\rm As}_2$ series,
a straightforward next question is, what will happen if Cu, the next $3d$, transition metal element, is doped into ${\rm BaFe}_2{\rm As}_2$?
Will the structural / magnetic phase transitions be suppressed in a similar manner? Will the superconducting dome shrink further? To answer these questions,
${\rm Ba(Fe}_{1-x}{\rm Cu}_x)_2{\rm As}_2$ single crystals were grown and characterized.
In Table I, we showed the results of the elemental analysis of the ${\rm Ba(Fe}_{1-x}{\rm Cu}_x)_2{\rm As}_2$ series.
We found Cu-doping has a somewhat larger variation of $x$ values than the other TM dopings (but still much less variation than K-doping).
This may come from the fact that small Cu shot rather
than CuAs powder was used in the growth procedure, but considering the fact that Co powder rather than CoAs powder
was used in reference \cite{chu} for the growth of ${\rm Ba(Fe}_{1-x}{\rm Co}_x)_2{\rm As}_2$ crystals and very sharp low field
$M(T)/H$ features were observed, it is more likely that this somewhat larger Cu-concentration variation is intrinsic in nature.

The evolution of the lattice parameters of ${\rm Ba(Fe}_{1-x}{\rm Cu}_x)_2{\rm As}_2$ with $x$ is shown in Fig. \ref{lcu}.
Comparing to pure ${\rm BaFe}_2{\rm As}_2$, with Cu doping up to $x$=0.356, the lattice parameter $a$ increases linearly
by 2.2\%, the lattice parameter
$c$ decreases monotonically by 1.7\% and the unit cell volume increases by roughly 2.6\%.

\bef
\psfig{file=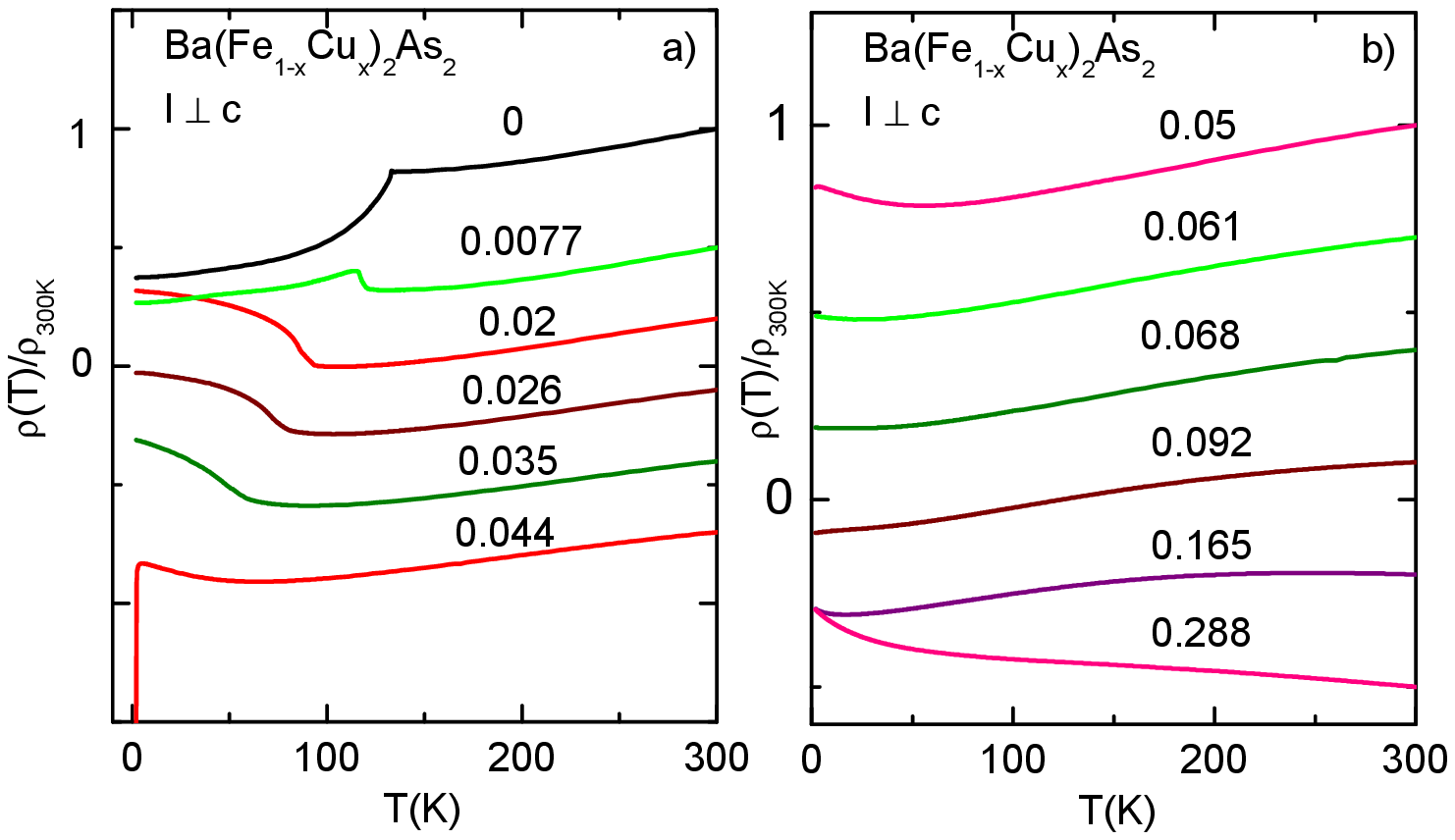,width=5in}
\caption{The temperature
dependent resistivity, normalized to the room temperature value, for
${\rm Ba(Fe}_{1-x}{\rm Cu}_x)_2{\rm As}_2$. Each subsequent data set is shifted downward by 0.3 for clarity respectively for (a) and (b).
}
\label{rcu}
\eef

\bef
\psfig{file=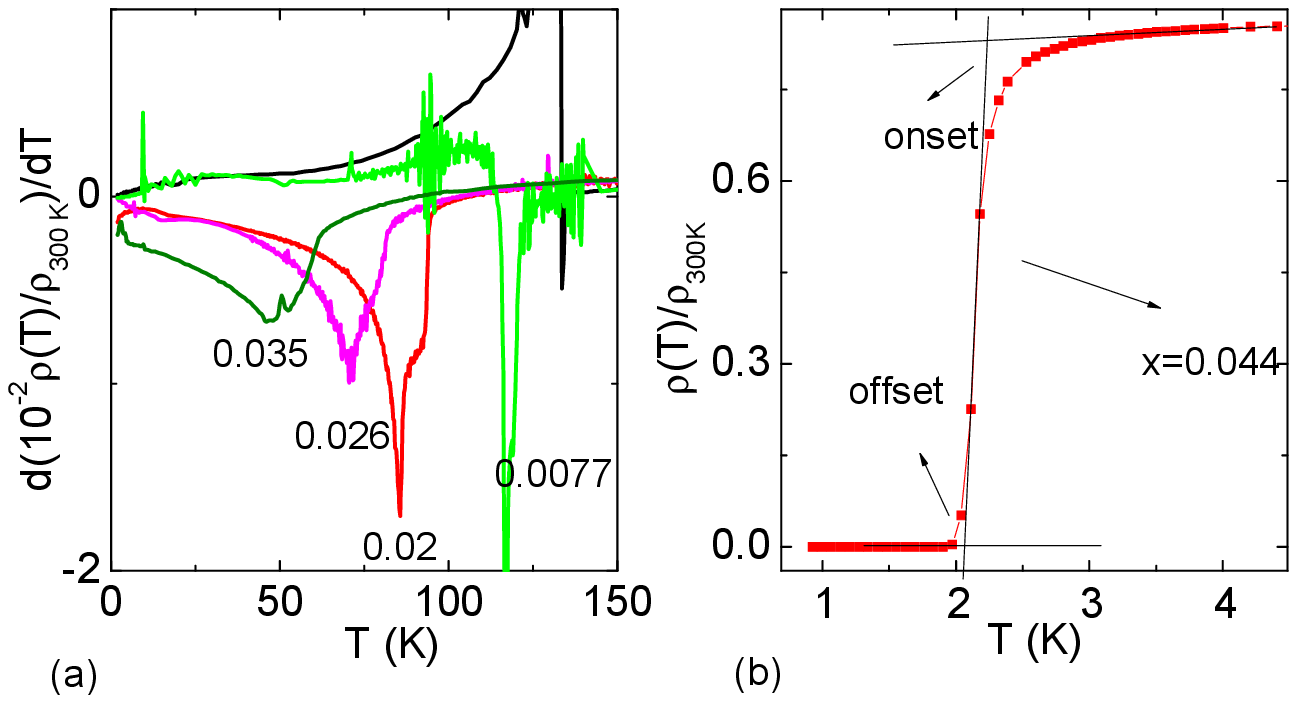,width=5in}
\caption{(a) $d(\rho(T)/\rho_{300K})/dT$ of ${\rm Ba(Fe}_{1-x}{\rm Cu}_x)_2{\rm As}_2$ for $0.035 \geq x$.
(b) Enlarged low temperature $\rho(T)/\rho_{300K}$ data of ${\rm Ba(Fe}_{0.956}{\rm Cu}_{0.044})_2{\rm As}_2$
}
\label{rcua}
\eef
The electrical transport data for the ${\rm Ba(Fe}_{1-x}{\rm Cu}_x)_2{\rm As}_2$ series from base temperature, 2 $K$, to 300 $K$ (for $x=0.044$, the
base temperature was 0.9 $K$) are shown in Fig. \ref{rcu};
the effects
of Cu substitution can be clearly seen. As $x$ is increased, the resistive anomaly
associated with the structural and magnetic phase transitions is
suppressed monotonically. For the lowest doping level, $x=0.0077$, the resistive anomaly manifests an abrupt increase
in resistivity similar to that found in pure ${\rm CaFe}_2{\rm As}_2$ \cite{nica} followed by a decrease as temperature is lowered further and
is very similar to what is shown in Fig. \ref{rni} for ${\rm Ba(Fe}_{0.9933}{\rm Ni}_{0.0067})_2{\rm As}_2$.
With higher Cu doping, the resistive anomalies associated with the structural and magnetic
phase transitions show a broad upturn. No clearly defined resistive anomaly can be seen
for $x>0.035$, but for $0.061\geq x >0.035$, a minimum in the resistivity can be observed, which can
be used to identify an upper limit for the structural and magnetic
phase transitions. No sign
 of structural and magnetic phase transitions is detected for $x\geq0.068$.
\bef
\psfig{file=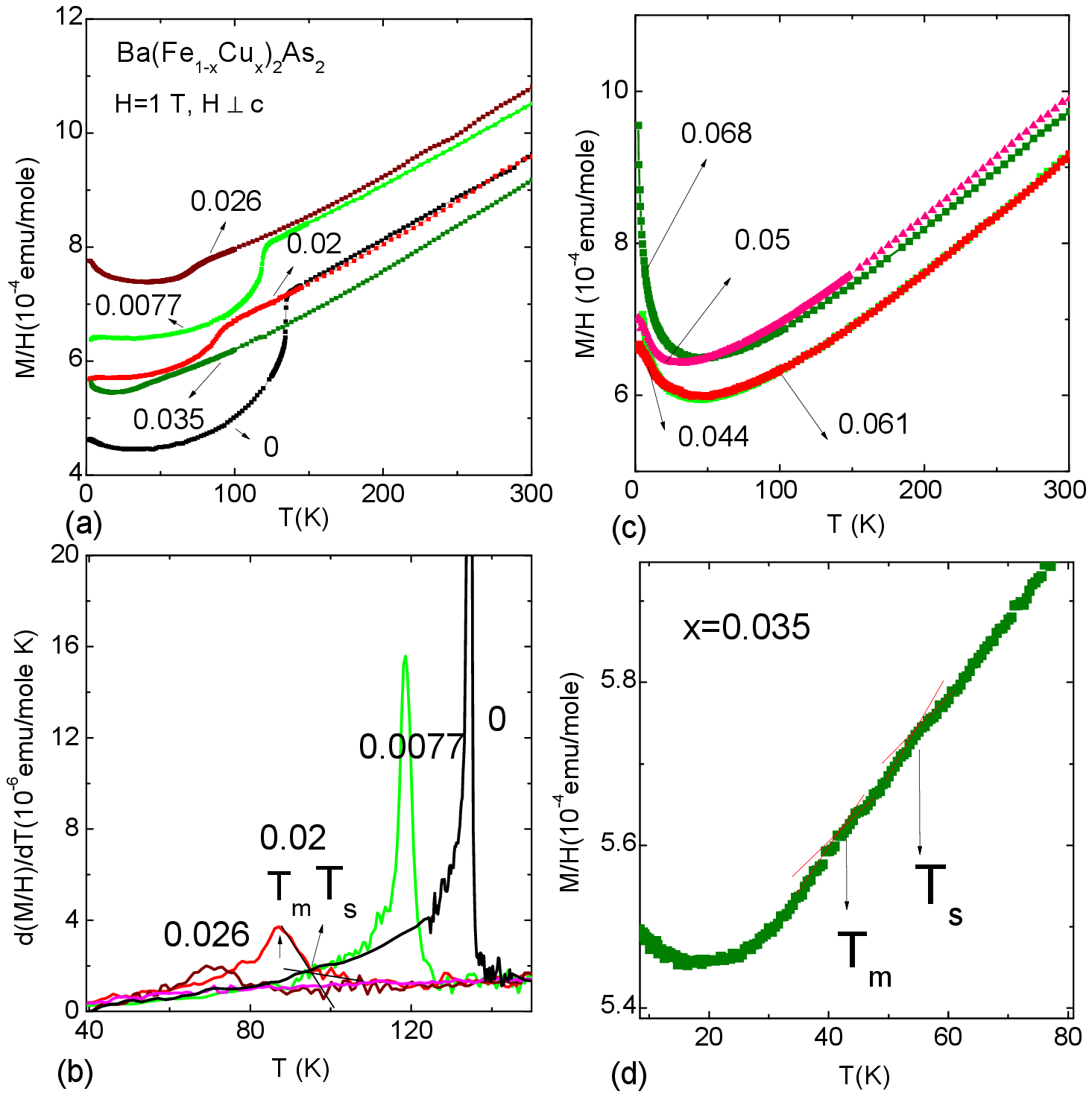,width=5in}
\caption{ The ${\rm Ba(Fe}_{1-x}{\rm Cu}_x)_2{\rm As}_2$ series: (a) $M(T)/H$ taken at 1 T with $H\bot c$ for $0\leq x \leq 0.035$.
(b) $M(T)/H$ taken at 1 T with $H\bot c$ for $0.035 < x \leq 0.068$.
(c) $d(M(T)/H)/dT$ for $x \leq 0.026$. The criteria to infer $T_s$ and $T_m$ are shown.
(d) The criteria to infer $T_s$ and $T_m$ for $x=0.035$ sample.
}
\label{mcu}
\eef
The suppression of the structural and magnetic phase transitions is further quantified in Fig. \ref{rcua}(a); two kinks, similar to
what we have seen in Co and Ni doped ${\rm BaFe}_2{\rm As}_2$ \cite{threedoping, nico, amesneutron},
can be observed. These features are suppressed to lower temperatures with increasing Cu doping.

Given the higher density and wider range of $x$-values studied in this work (as compared to reference \cite{threedoping}),
zero resistivity was found for a single doping: $x=0.044$, below 2.1 K.
Figure \ref{rcua}(b) shows the enlarged, low temperature, electric transport data of ${\rm Ba(Fe}_{0.956}{\rm Cu}_{0.044})_2{\rm As}_2$.
A very sharp transition to zero resistivity is observed. $T_c^{offset}$ is 2.1 K and $T_c^{onset}$ is 2.2 K.

Figure \ref{mcu} shows the temperature dependent $M(T)/H$ data taken at 1 T from 2 K to 300 K with $H$ perpendicular
to the crystallographic $c$-axis of the samples. Due to slight ferromagnetic impurities in the
higher Cu concentration ${\rm BaFe}_2{\rm As}_2$ samples ($x>0.068$), we only show the susceptibility for $x\leq0.068$.
To make the graphs easier to read, the data are grouped
into two sets. Figure \ref{mcu} (a) shows $M(T)/H$ for ${\rm Ba(Fe}_{1-x}{\rm Cu}_x)_2{\rm As}_2$ samples that manifest
magnetic anomalies in Fig. \ref{rcu}.
  A clear drop at the temperature associated with the magnetic anomalies can be seen.
     Higher temperature susceptibility data show
  essentially linear temperature dependence, similar to the ${\rm Ba(Fe}_{1-x}{\rm Ni}_x)_2{\rm As}_2$ data shown in Fig. \ref{mni}.
 Figure \ref{mcu} (b) shows $M(T)/H$ for ${\rm Ba(Fe}_{1-x}{\rm Cu}_x)_2{\rm As}_2$ samples ($0.068 \geq x > 0.035$).
 Although a resistivity minimum is present in $x=0.05, 0.061$ and 0.068 samples, no clear
 feature of structural or magnetic phase transitions, similar to
 the ones in Fig. \ref{mcu} (a), can be detected in the susceptibility data. On the other hand,
 the low temperature susceptibility increases with decreasing
 temperature whereas the high temperature susceptibility retains its almost linear-in-T behavior.
Low field $M(T)/H$ data, down to 1.8 K, for the $x$=0.044 sample, with zero resistivity around 2 K,
do not show a diamagnetic signal,
 but since this is at the edge of range where diamagnetic signal would just be starting, it is hard to conclude if
there is (or isn't) bulk superconductivity in $x$=0.044 sample. To infer $T_s$ and $T_m$,
$d(M/H)/dT$ are plotted in Fig. \ref{mcu} (c). Due to the blurred features in $d(M/H)/dT$, the criteria to infer $T_s$
for Cu-doping series are different from Ni-doping series, as shown in Fig. \ref{mcu} (c). Figure \ref{mcu} (d) shows the manner to infer
$T_s$ and $T_m$ for $x=0.035$ sample.

\bef
\psfig{file=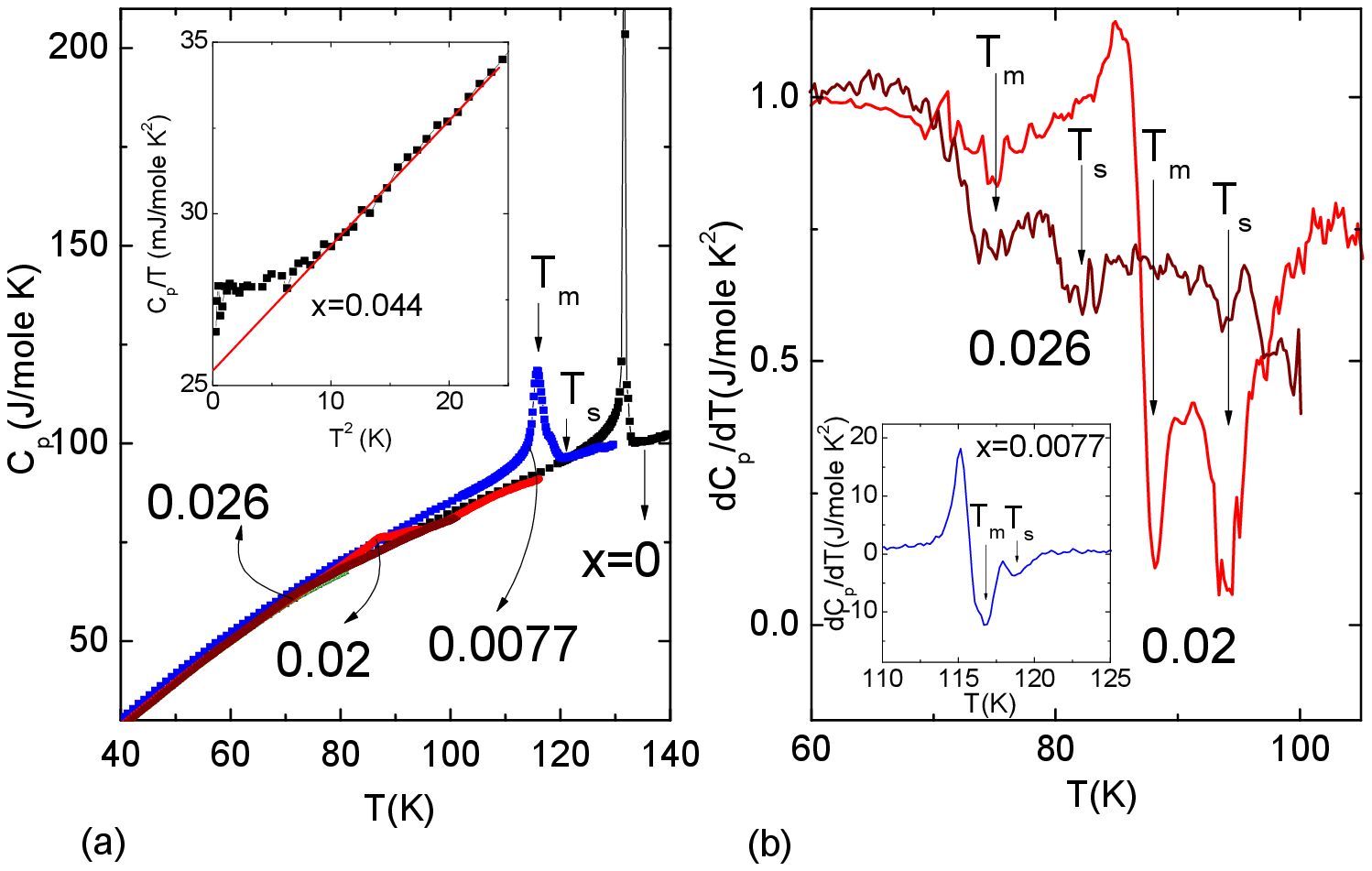,width=5in}
\caption{(a) Temperature dependent heat capacity of ${\rm Ba(Fe}_{1-x}{\rm Cu}_{x})_2{\rm As}_2$ ($x=0$, 0.0077, 0.02 and 0.026).
Inset: $C_p$ vs. $T^2$ for ${\rm Ba(Fe}_{0.956}{\rm Cu}_{0.044})_2{\rm As}_2$.
(b) $dC_p/dT$ vs. T for ${\rm Ba(Fe}_{1-x}{\rm Cu}_{x})_2{\rm As}_2$ ($x=$ 0.0077, 0.02 and 0.026).}
\label{ccu}
\eef
Figure \ref{ccu} (a) shows the specific heat data $C_p(T)$ for the Cu
concentrations $x=0$, 0.0077, 0.02 and 0.026 for temperature near the structural and magnetic phase transitions. The very sharp peak around 134 K
associated with the structural / magnetic phase transition can be seen in heat capacity measurement for
pure ${\rm BaFe}_2{\rm As}_2$.
For $x=0.0077$, the single sharp peak in pure ${\rm BaFe}_2{\rm As}_2$ splits into two features in $dC_p/dT$
as seen in the inset of Fig. \ref{ccu} (b).
With even higher Cu dopings, the sharp peaks become broad.
To identify these features more clearly, $dC_p/dT$ for $x$ = 0.02 and 0.026 are plotted in Fig. \ref{ccu} (b). We can see two kinks in the
$dC_p/dT$ plot which correspond to the two kinks observed in $d(\rho/\rho_{300})/dT$ \cite{nico}.
These features are no longer detectable
in either $C_p$ or $dC_p/dT$ for $x\geq 0.035$.
The inset of Fig. \ref{ccu} (a) shows the $C_p$ vs. $T^2$ measured down to 0.9 K for ${\rm Ba(Fe}_{0.956}{\rm Cu}_{0.044})_2{\rm As}_2$.
Although there is a clear break from the linear behavior
seen for $T^2 < 4 K^2$, no sharp jump associated with superconductivity can be observed around 4 $K^2$. This is not surprising since the heat capacity
jump decreases with decreasing $T_c$ \cite{sergey}: for Co-doped and Ni-doped ${\rm BaFe}_2{\rm As}_2$,
 the heat capacity jump is rather subtle for superconductors with very low $T_c$ values due to the broadness, such as
  Ni doped ${\rm BaFe}_2{\rm As}_2$ samples with $T_c$ around
2.5 K and 4 K, neither of which showed a clear specific heat jump.
\bet
\begin{tabular}{c | c | c c c c| c c | c c}
   \hline
   \hline

    dopant & $x$ & \multicolumn{4}{|c|}{$\rho$}& \multicolumn{2}{|c|}{$M$}& \multicolumn{2}{c}{$C$}\\
    \hline
   Cu&~                & {$T_s$}& {$T_m$} & {$T_c^{onset}$} & {$T_c^{offset}$}& {$T_s$} & {$T_m$} & {$T_s$}& {$T_m$}\\

  ~&         0.0077  & 119  & 117    &                &             &    &119    & 119   &117  \\

     ~&      0.02   & 93    &86      &                 &           &96    &88     &  94     &88   \\

     ~&      0.026  &79     &71      &                  &           &78    & 72    &  82     &75  \\

     ~&     0.035  &57    &48      &                 &            &56    & 42         &        &    \\

      ~&     $0.044$  &$40\pm20^{**}$     &      & 2.2             & 2.1     &          &         &        &    \\

      ~&     $0.05$   &$30^\pm25^{**}$     &          &                 &        &        &          &        &    \\

      ~&     $0.061$  &$10\pm10^{**}$     &          &                 &        &        &          &        &       \\
\hline
   \hline
  \end{tabular}
\caption{Summary of $T_s$, $T_m$ and $T_c$ from resistivity, magnetization and specific heat measurements for the
 ${\rm Ba(Fe}_{1-x}{\rm Cu}_x)_2{\rm As}_2$ series. *: see text.}
  \label{tablecu}
\eet

\bef
\psfig{file=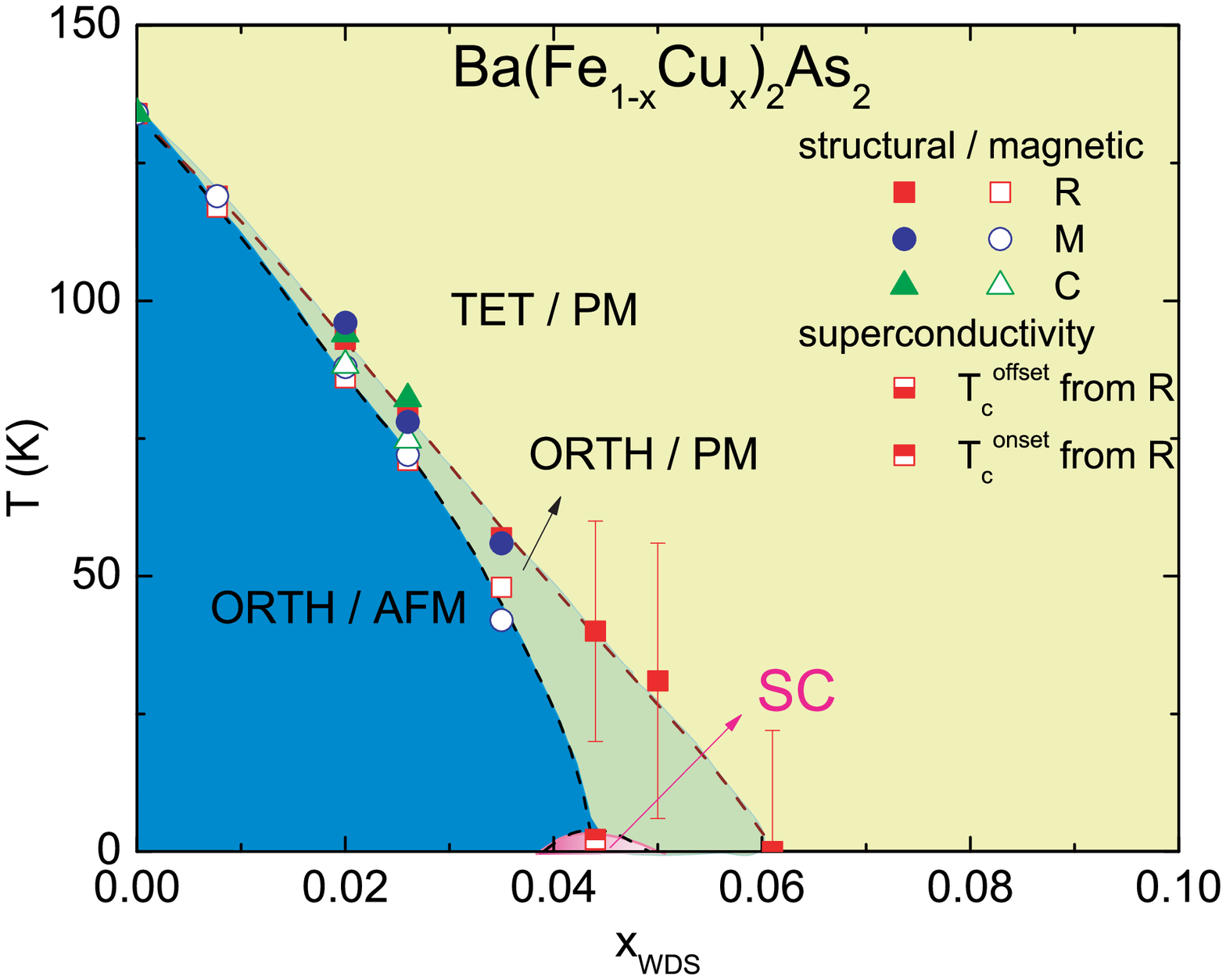,width=4in}
\caption{$T - x$ phase diagram of
Ba(Fe$_{1-x}$Cu$_x$)$_2$As$_2$ single crystals for $x \leq 0.061$.
Superconductivity is only determined below ~ 2 K, the extent of the superconducting
region is unknown, but is bounded by $x=0.035$ on the underdoped side and
$x=0.05$ on the overdoped side.
}\label{phasecu}
\eef

The structural / magnetic and superconducting transition temperatures are determined
from Figs. \ref{rcu} - \ref{ccu} and summarized in Table \ref{tablecu} and Fig. \ref{phasecu}.
For the data indexed by $**$, the resistive
features have become so broad that the error bars associated with the
determination of the upper (only detectable) transition are
defined by the temperature of the resistance minima on the high
side and the temperature of the inflection point on the low side.
The $T-x$ phase diagram of the ${\rm Ba(Fe}_{1-x}{\rm Cu}_x)_2{\rm As}_2$ series is plotted in Fig. \ref{phasecu}.
The structural and magnetic phase transitions are suppressed and increasingly split with Cu doping in a similar manner as Co, Ni dopings,
but superconductivity is only detected for $x=$0.044,
with a very low $T_c$ ($\sim 2$ K).
Given the narrow range of superconductivity,
the extent of the superconducting dome and how $T_m$ intersects it (if
indeed it does) are speculation.

\subsection{${\rm Ba(Fe}_{1-x-y}{\rm Co}_x{\rm Cu}_y)_2{\rm As}_2$ ($x \sim 0.022$)}

\bef \psfig{file=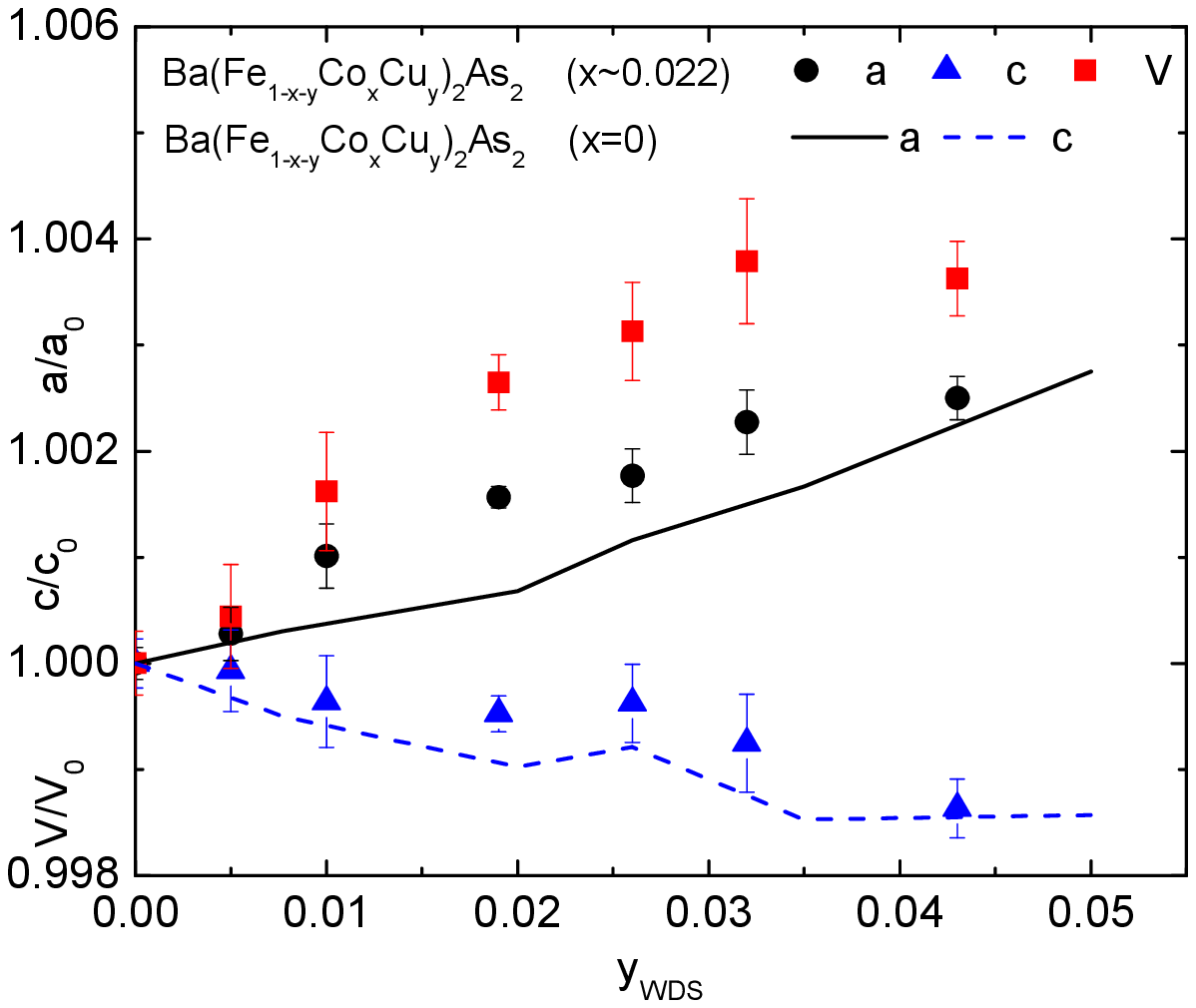,width=4in} \caption{
Lattice parameters of the ${\rm Ba(Fe}_{1-x-y}{\rm Co}_x{\rm Cu}_y)_2{\rm As}_2$ ($x \sim 0.022$) series,
$a$ and $c$ as well as unit cell volume, $V$, normalized to
 the values of ${\rm Ba(Fe}_{0.976}{\rm Co}_{0.024})_2{\rm As}_2$ ($a_0$=3.9598(6)$\AA$, $c_0$=13.004(3)$\AA$) as
 a function of measured Cu concentration, $y_{WDS}$. The solid lines represent
 the values of $a/a_0$ and $c/c_0$ for the ${\rm Ba(Fe}_{1-x-y}{\rm Co}_x{\rm Cu}_y)_2{\rm As}_2$ ($x=0$) series shown in Fig. 9.
} \label{lcocua} \eef

Whereas doping ${\rm BaFe}_2{\rm As}_2$ with Co, Ni or Cu suppresses the upper structural / magnetic phase transitions in similar ways,
only Co and Ni appear
 to induce a superconducting dome over substantial ranges of $x$ values. Cu, while suppressing
  the structural and magnetic phase transitions, does not lead to a
  significant superconducting region; so far only one compound with
 $x\sim 0.044$ has $T_c\sim 2$ K. In order to better understand the effects of Cu on the superconducting state, two mixed (Cu and Co) doping series,
  ${\rm Ba(Fe}_{1-x-y}{\rm Co}_x{\rm Cu}_y)_2{\rm As}_2$ ($x \sim 0.022$ and $x \sim 0.047$) were grown and studied.

For the ${\rm Ba(Fe}_{1-x-y}{\rm Co}_x{\rm Cu}_y)_2{\rm As}_2$ ($x \sim 0.022$) series,
the lattice parameters are normalized to the ones of the closely related ${\rm Ba(Fe}_{0.976}{\rm Co}_{0.024})_2{\rm As}_2$.
$a/a_0$, $c/c_0$ and $V/V_0$ are plotted against $y_{WDS}$ in Fig. \ref{lcocua}. With Cu doped into
${\rm Ba(Fe}_{0.978}{\rm Co}_{0.022})_2{\rm As}_2$, the lattice parameter $a$ increases and the
lattice parameter $c$ decreases. These changes are in qualitatively similar manners to the ones
when Cu was doped into ${\rm BaFe}_2{\rm As}_2$ (Fig. \ref{lcu}), which are presented
in Fig. \ref{lcocua} as solid lines.

\bef
\psfig{file=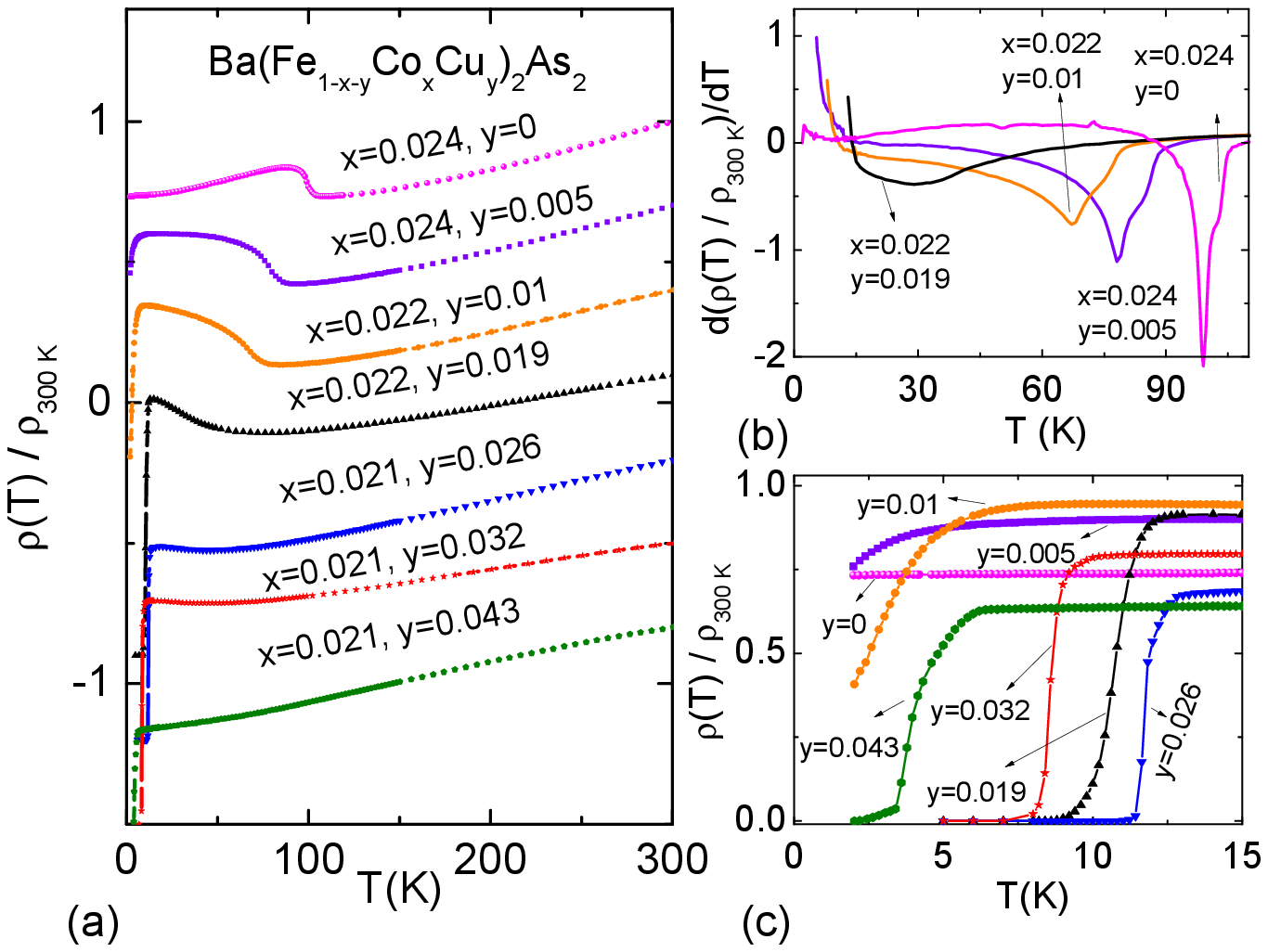,width=5in}
\caption{The ${\rm Ba(Fe}_{1-x-y}{\rm Co}_x{\rm Cu}_y)_2{\rm As}_2$ ($x \sim 0.022$) series.
(a) The temperature
dependent resistivity, normalized to the room temperature values. Each subsequent data set is shifted downward by 0.3 for clarity.
(b) $d(\rho(T)/\rho_{300K})/dT$ for $y \leq 0.019$.
(c) Enlarged, low temperature, $\rho(T)/\rho_{300K}$.
}
\label{rcocua}
\eef

Figure \ref{rcocua} (a) shows the electric transport data for the ${\rm Ba(Fe}_{1-x-y}{\rm Co}_x{\rm Cu}_y)_2{\rm As}_2$ ($x\sim 0.022$)
series from 2 K to 300 K.
For ${\rm Ba(Fe}_{0.976}{\rm Co}_{0.024})_2{\rm As}_2$ ($y=0$), no sign of superconductivity can be detected;
as the temperature is reduced from 300 K, the resistivity
exhibits an upturn around 110 K and then decreases with further cooling.
When Cu is doped into ${\rm Ba(Fe}_{0.978}{\rm Co}_{0.022})_2{\rm As}_2$, the structural / magnetic phase transitions are
suppressed to lower temperature
and evolve in a manner that is qualitatively similar to what is found for other $TM$ dopings.
Figure \ref{rcocua} (b) shows the
derivative of $\rho(T)/\rho_{300K}$. Similar to Co, Ni and Cu doped ${\rm BaFe}_2{\rm As}_2$, two kinks are seen
to separate and suppressed to lower temperature as more Cu is added.
For intermediate $y$ values, superconductivity can be stabilized. Figure \ref{rcocua} (c) shows an expanded
plot of the low temperature, $\rho(T)/\rho_{300K}$ data.
When $y=0.019$, zero resistivity is detected below 9 K.
$T_c$ reaches a maximum of 12 K for $y=0.026$ and drops to 8.3 K for $y=0.032$ and 2 K for $y=0.043$.

\bef
\psfig{file=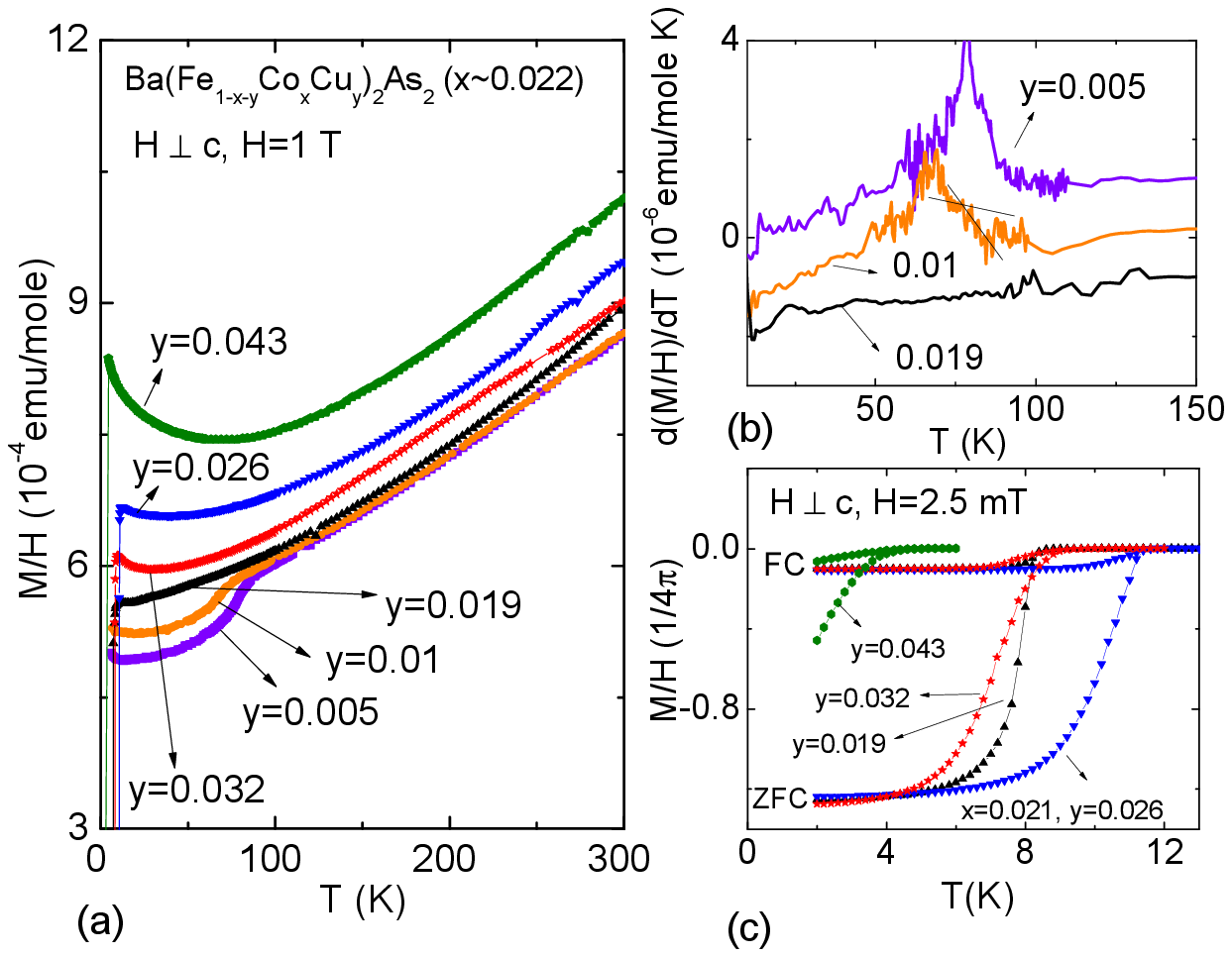,width=5in}
\caption{The ${\rm Ba(Fe}_{1-x-y}{\rm Co}_x{\rm Cu}_y)_2{\rm As}_2$ ($x \sim 0.022$) series:
(a) $M(T)/H$ data taken at 1 T with $H\bot c$.
(b) $d(M/H)/dT$ for $y=0.005$, 0.01 and 0.019 samples. Each subsequent data set is shifted downward by $1\times 10^{-6}$ for clarity.
(c) Field-cooled (FC) and zero-field-cooled (ZFC) low field $M(T)/H$ data taken at 2.5 mT with $H\bot c$.
}
\label{mcucoa}
\eef

Figure \ref{mcucoa} (a) shows the $ M(T)/H$
data taken at 1 T with H perpendicular to the
crystallographic $c$-axis of the ${\rm Ba(Fe}_{1-x-y}{\rm Co}_x{\rm Cu}_y)_2{\rm As}_2$ series. The high temperature drop in the susceptibility
data is associated with the structural / magnetic phase transitions, and consistent
with the resistivity measurements. The high temperature close-to-linear
susceptibility can also be seen in this series. The magnitude of the susceptibility is comparable to those of ${\rm Ba(Fe}_{1-x}{\rm Co}_x)_2{\rm As}_2$
and ${\rm Ba(Fe}_{1-x}{\rm Ni}_x)_2{\rm As}_2$. $d(M/H)/dT$ is plotted in Fig. \ref{mcucoa} (b). Two kinks can be seen
for $y=0.005$ and 0.01 samples.

Figure \ref{mcucoa} (c) shows the ${M(T)/H}$ data taken at 2.5 mT with H perpendicular to the
crystallographic $c$-axis of the ${\rm Ba(Fe}_{1-x-y}{\rm Co}_x{\rm Cu}_y)_2{\rm As}_2$ ($x \sim 0.022$) series.
Superconductivity can be clearly seen in the field-cooled (FC) and zero-field-cooled (ZFC) data.
Comparing the low field ${M(T)/H}$ data with the ones in ${\rm Ba(Fe}_{1-x}{\rm Co}_x)_2{\rm As}_2$ \cite{nico}, we can see that these two series have
very similar superconducting volume fractions / pinning. It is worth noting that, as a "reality check", since
the superconductivity in the ${\rm Ba(Fe}_{1-x-y}{\rm Co}_x{\rm Cu}_y)_2{\rm As}_2$ series has a superconducting volume that is comparable to that
of the ${\rm Ba(Fe}_{1-x}{\rm Co}_x)_2{\rm As}_2$ phase, superconductivity must come from a bulk phase.
\bef
\psfig{file=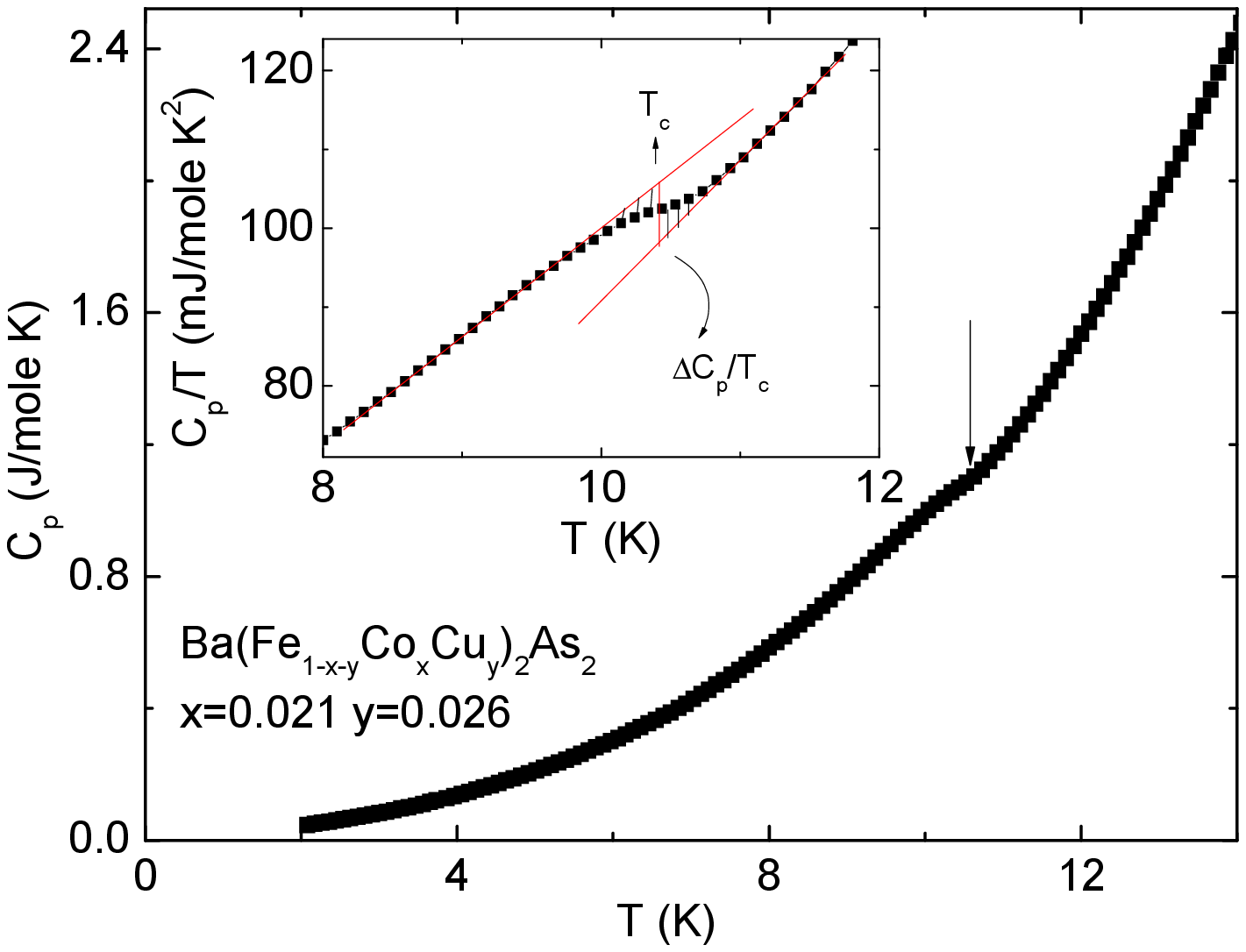,width=4in}
\caption{Temperature dependent heat capacity of ${\rm Ba(Fe}_{0.953}{\rm Co}_{0.021}{\rm Cu}_{0.026})_2{\rm As}_2$.
Inset: $C_p$/T vs. T.}
\label{ckq158}
\eef
The width of the superconducting transition shown in Fig. \ref{mcucoa} (c) is not quite as sharp as that found for the
higher $x$-value ${\rm Ba(Fe}_{1-x}{\rm Co}_x)_2{\rm As}_2$ samples,
 this could imply that the ${\rm Ba(Fe}_{1-x-y}{\rm Co}_x{\rm Cu}_y)_2{\rm As}_2$ samples are not
 as homogeneous as ${\rm Ba(Fe}_{1-x}{\rm Co}_x)_2{\rm As}_2$ ones.
This is consistent with the WDS measurements, summarized in Table I, which, although showing a homogeneous Co concentration for the
${\rm Ba(Fe}_{1-x}{\rm Co}_x)_2{\rm As}_2$ and ${\rm Ba(Fe}_{1-x-y}{\rm Co}_x{\rm Cu}_y)_2{\rm As}_2$
series, indicates that the Cu concentration has a variation of up to 10\% of the real Cu concentration in
both the ${\rm Ba(Fe}_{1-x}{\rm Cu}_x)_2{\rm As}_2$ and
${\rm Ba(Fe}_{1-x-y}{\rm Co}_x{\rm Cu}_y)_2{\rm As}_2$ series.

Heat capacity data was collected for ${\rm Ba(Fe}_{0.953}{\rm Co}_{0.021}{\rm Cu}_{0.026})_2{\rm As}_2$,
the composition that manifests the highest $T_c$ in this series.
A clear heat capacity jump can be seen in Fig. \ref{ckq158} around 11 K. The inset shows temperature dependent $C_p/T$ data near $T_c$. $T_c$ and
$\triangle C_p/T_c$ were inferred using an "isoentropic" construction \cite{sergey} so that the two areas shown in Fig. \ref{ckq158} have equal areas;
$\triangle C_p/T_c$ is 7.6 mJ/mole $K^2$ with $T_c$ equal to 10.4 K.
These values fall onto the $log(\triangle C_p/T_c)$ vs. $logT$ plot shown in reference \cite{sergey}.

From Figs. \ref{rcocua}, \ref{mcucoa} and \ref{ckq158}, we can determine
the structural / magnetic and superconducting transition temperatures for the ${\rm Ba(Fe}_{1-x-y}{\rm Co}_x{\rm Cu}_y)_2{\rm As}_2$ ($x \sim 0.022$)
series. These results are summarized in Table \ref{tablecocua} and graphically presented as a $T-y$ phase diagram in Fig. \ref{phasecocua}.
For the temperature indexed by $**$, $T_s$ was inferred via the same way as we infer $T_s$ for
the temperatures indexed by $**$ in the ${\rm Ba(Fe}_{1-x}{\rm Cu}_x)_2{\rm As}_2$ series.
For the temperature indexed by $*$, the criteria in the inset of Fig. \ref{rni} (b) are employed.

 Figure \ref{phasecocua} shows that the structural and magnetic phase transitions are suppressed and increasingly split with doping,
in addition, superconductivity is stabilized in a dome-like region. The phase diagram has a very similar appearance to those found
for the Co-doped and Ni-doped series.

\bet
\begin{tabular}{c | c | c | c c c c| c c c | c}
   \hline
   \hline

    dopant & $x$ & $y$  &\multicolumn{4}{|c|}{$\rho$}& \multicolumn{3}{|c|}{$M$}& \multicolumn{1}{c}{$C$}\\
    \hline
   Cu / Co& ~&~                      & {$T_s$}& {$T_m$}& {$T_c^{onset}$}& {$T_c^{offset}$} &  {$T_s$}& {$T_m$}& {$T_c$} & {$T_c$}\\

  ~&      0.024 &0          &103     & 99       &   &                             &     &             &  &      \\

  ~&      0.024 &0.005      &85      & 78      &    &                             & 85    &79     &  &      \\

   ~&     0.022 &0.01       &75     & 68   &4.7 &0                            &78  &68        &  &    \\

  ~ &  0.022 &0.019      & 41        &$29^{*}$      &  11  &9                            &       &           & 8.7   &   \\

     ~&   0.021 &$0.026$      &$25\pm 15^{**}$           &        &  12.1     &11                       &         &           &11.6  &10.4  \\

      ~&  0.021 &0.032      &        &        &  8.9     &8.3                         &         &         &9.6  &    \\

      ~&  0.021 &0.043     &         &        &  4.6     &2                              &      &       &4.3   &      \\
\hline
   \hline

  \end{tabular}

\caption{Summary of $T_s$, $T_m$ and $T_c$ from resistivity, magnetization and specific heat measurements for the
${\rm Ba(Fe}_{1-x-y}{\rm Co}_x{\rm Cu}_y)_2{\rm As}_2$ ($x \sim 0.022$) series. * and **: see text.}
\label{tablecocua}
\eet
\bef
\psfig{file=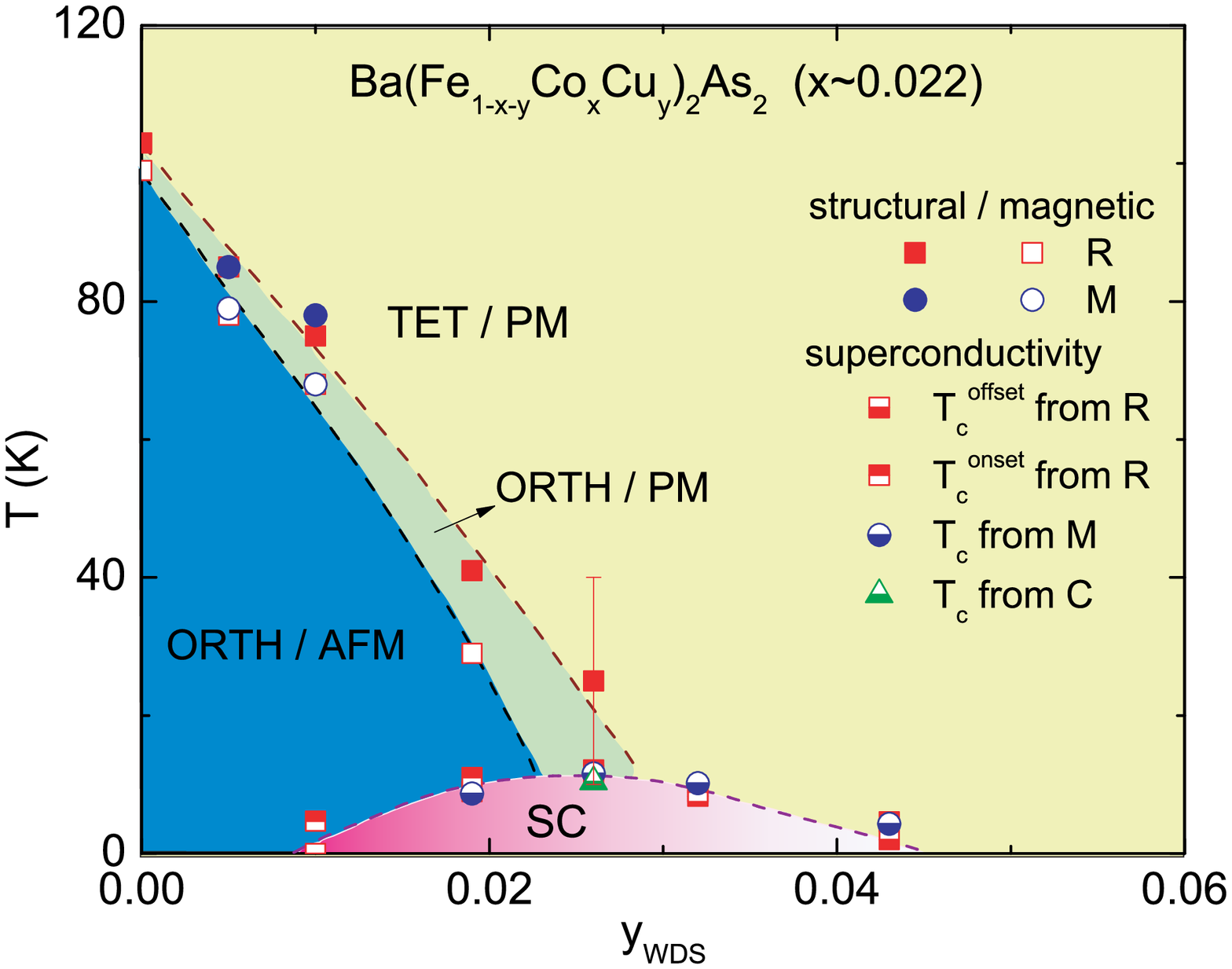,width=4in}
\caption{$T - y$ phase diagram of
${\rm Ba(Fe}_{1-x-y}{\rm Co}_x{\rm Cu}_y)_2{\rm As}_2$ ($x \sim 0.022$) single crystals.
The shading in the superconducting dome implies the existence of a crossover
from tetragonal / paramagnetic phase to orthorhombic / antiferromagnetic phase,
as used in Fig. \ref{phaseni}.
}\label{phasecocua}
\eef

\bef
\psfig{file=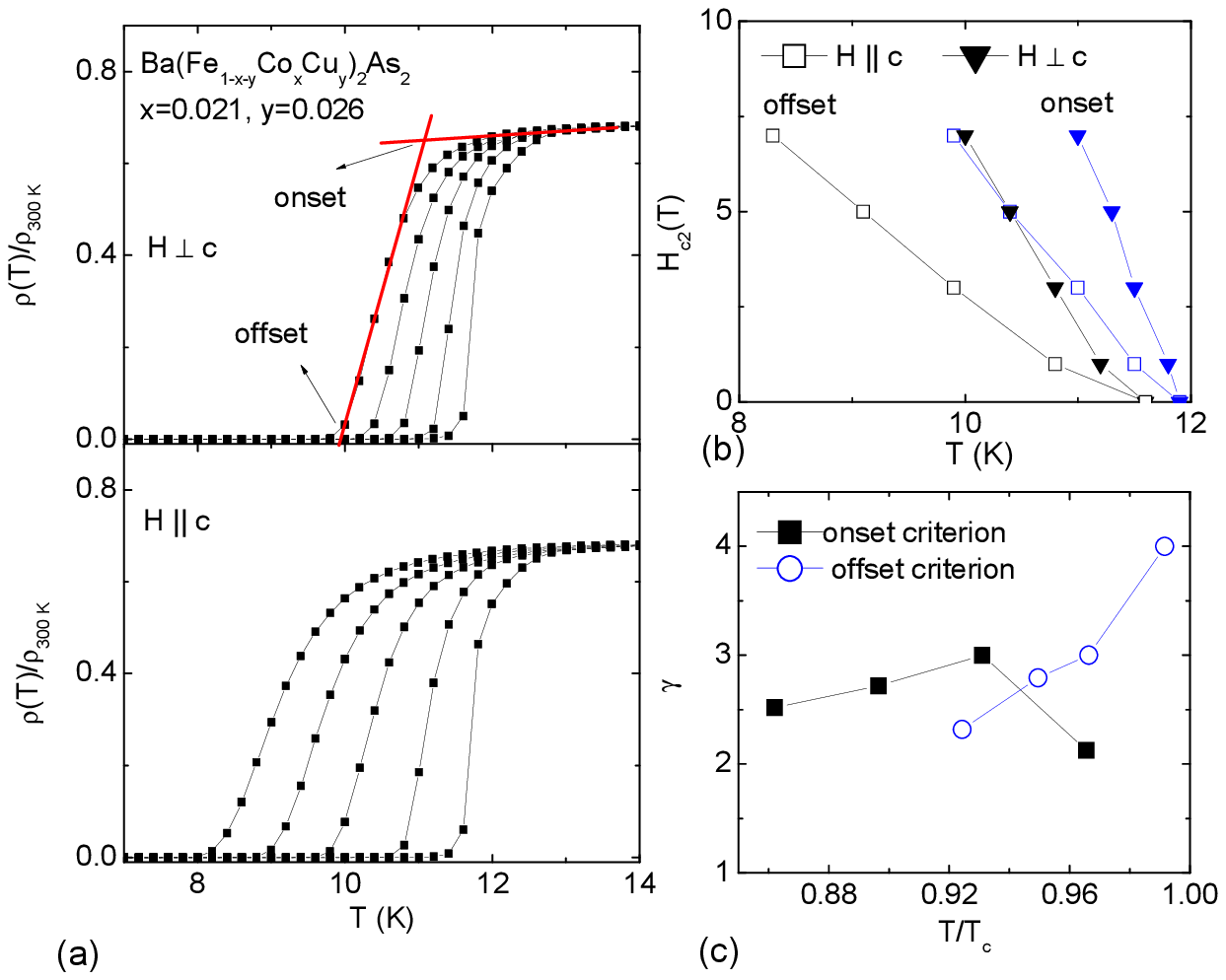,width=5in}
\caption{(a)Low temperature
dependent resistivity, normalized by room temperature value, for
${\rm Ba(Fe}_{0.953}{\rm Co}_{0.021}{\rm Cu}_{0.026})_2{\rm As}_2$
 with 0 T, 1 T, 3 T, 5 T and 7 T magnetic field perpendicular to $c$-axis (upper panel)
and along $c$-axis(lower panel). (b)Critical field $H_{c2}$ vs. T determined from onset and offset criteria.
(c) The ratio of anisotropic critical field $\gamma=H_{c2}^{\bot c}/H_{c2}^{||c}$ vs. $T/T_c$.
}
\label{hc2kq158}
\eef

Figure \ref{hc2kq158} (a) shows the low temperature
$\rho(T)/\rho_{300K}$ vs. T data taken at 0 T, 1 T, 3 T, 5 T and 7 T for ${\rm Ba(Fe}_{0.953}{\rm Co}_{0.021}{\rm Cu}_{0.026})_2{\rm As}_2$ when
H is applied perpendicular to the $c$-axis and along the $c$-axis, respectively. $T_c$ decreases
with increasing applied magnetic field more rapidly for $H || c$. The offset and onset criteria used
to infer $T_c$ are shown in Fig. \ref{hc2kq158} (a).
The temperature-dependent, resistive $H_{c2}(T)$ curves are plotted in Fig. \ref{hc2kq158} (b).
Using onset criterion, $(dH_{c2}^{\bot c}/dT)|_{T_c}$ is about -7.8 T/K and $(dH_{c2}^{||c}/dT)|_{T_c}$ is about -3.5 T/K.
Using offset criterion, $(dH_{c2}^{\bot c}/dT)|_{T_c}$ is about -4.5 T/K and $(dH_{c2}^{||c}/dT)|_{T_c}$ is about -2.2 T/K.
Using the WHH equation, $H_{c2}^{\bot}(0)$ is estimated
to be 64 T for onset criterion and 36 T for offset criterion and $H_{c2}^{||c}(0)$ is
to be 28 T for onset criterion and 18 T for offset criterion. Figure \ref{hc2kq158} (c) shows the
anisotropy of the upper critical field $\gamma$=$H_{c2}^{\bot}/H_{c2}^{||c}$, which was calculated in the same manner as outlined
for Fig. \ref{hc2ni1}. As we can see, in the range of $T/T_c$ from 0.85 to 0.99,
$\gamma$ varies between 2 to 3 for onset criterion and 2 to 4 for offset
criterion, which is comparable to the optimal and over-doped
${\rm Ba(Fe}_{1-x}{\rm Co}_x)_2{\rm As}_2$ \cite{nico} and the optimally doped ${\rm Ba(Fe}_{1-x}{\rm Ni}_x)_2{\rm As}_2$ (this work).

\subsection{${\rm Ba(Fe}_{1-x-y}{\rm Co}_x{\rm Cu}_y)_2{\rm As}_2$ ($x \sim 0.047$)}
It is worth noting that the maximum $T_c$ value for the ${\rm Ba(Fe}_{1-x-y}{\rm Co}_x{\rm Cu}_y)_2{\rm As}_2$ ($x \sim 0.022$) series is around 12 K,
which is somewhat low in comparison to the Co- or Ni- doped series. To study the effects of Cu doping further,
a {${\rm Ba(Fe}_{1-x-y}{\rm Co}_x{\rm Cu}_y)_2{\rm As}_2$ ($x \sim 0.047$)} series was grown and examined. For $y=0$, this is an underdoped,
  but superconducting, member of the ${\rm Ba(Fe}_{1-x}{\rm Co}_x)_2{\rm As}_2$ series. The elemental analysis shown in Table I
  indicates that within a single batch the variation of Cu concentration is roughly $\pm$10\% of the average concentration, similar to
  the variation range in the ${\rm Ba(Fe}_{1-x}{\rm Cu}_x)_2{\rm As}_2$ and
 ${\rm Ba(Fe}_{1-x-y}{\rm Co}_x{\rm Cu}_y)_2{\rm As}_2$ ($x \sim 0.022$) series.

\bef
\psfig{file=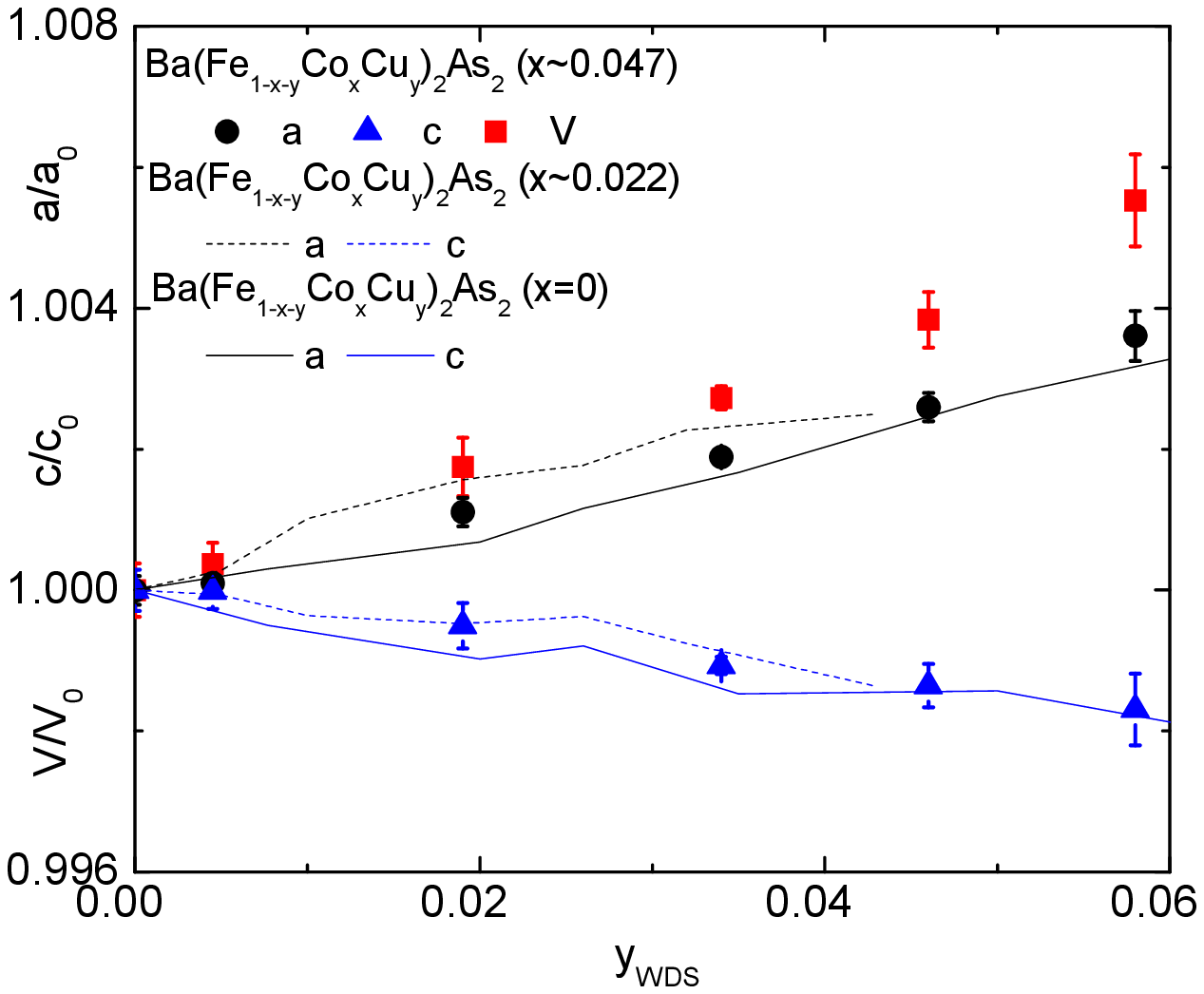,width=4in} \caption{
Lattice parameters of the ${\rm Ba(Fe}_{1-x-y}{\rm Co}_x{\rm Cu}_y)_2{\rm As}_2$ ($x \sim 0.047$) series,
$a$ and $c$ as well as unit cell volume, $V$, normalized to
 the values of ${\rm Ba(Fe}_{0.953}{\rm Co}_{0.047})_2{\rm As}_2$ ( $a_0$=3.9605(6)$\AA$, $c_0$=12.992(4)$\AA$) as
 a function of measured Cu concentration, $y_{WDS}$. The dash lines and solid lines represent
 the values for the ${\rm Ba(Fe}_{1-x-y}{\rm Co}_x{\rm Cu}_y)_2{\rm As}_2$ ($x \sim 0.022$) and
 ${\rm Ba(Fe}_{1-x-y}{\rm Co}_x{\rm Cu}_y)_2{\rm As}_2$ ($x=0$) series shown in Fig. 9 and Fig. 15 respectively.
}
\label{lcocub}
\eef

Figure \ref{lcocub} presents the normalized lattice parameters $a/a_0$, $c/c_0$ and $V/V_0$ for
this series, where $a_0$, $c_0$
and $V_0$ are the ones for ${\rm Ba(Fe}_{0.953}{\rm Co}_{0.047})_2{\rm As}_2$. As Cu is doped into
${\rm Ba(Fe}_{0.953}{\rm Co}_{0.047})_2{\rm As}_2$, the lattice parameter $a$ and unit cell volume increase while the
lattice parameter $c$ decreases. As a comparison, the curves of
$a/a_0$ and $c/c_0$ of the ${\rm Ba(Fe}_{1-x-y}{\rm Co}_x{\rm Cu}_y)_2{\rm As}_2$ (x=0, $x \sim 0.022$)
series presented in Fig. 9 and Fig. 15 are added as dash line and solid line in Fig. 21. As we can see, the effects of Cu doping on
the lattice parameters of these series, are quantitatively similar to each other.

\bef
\psfig{file=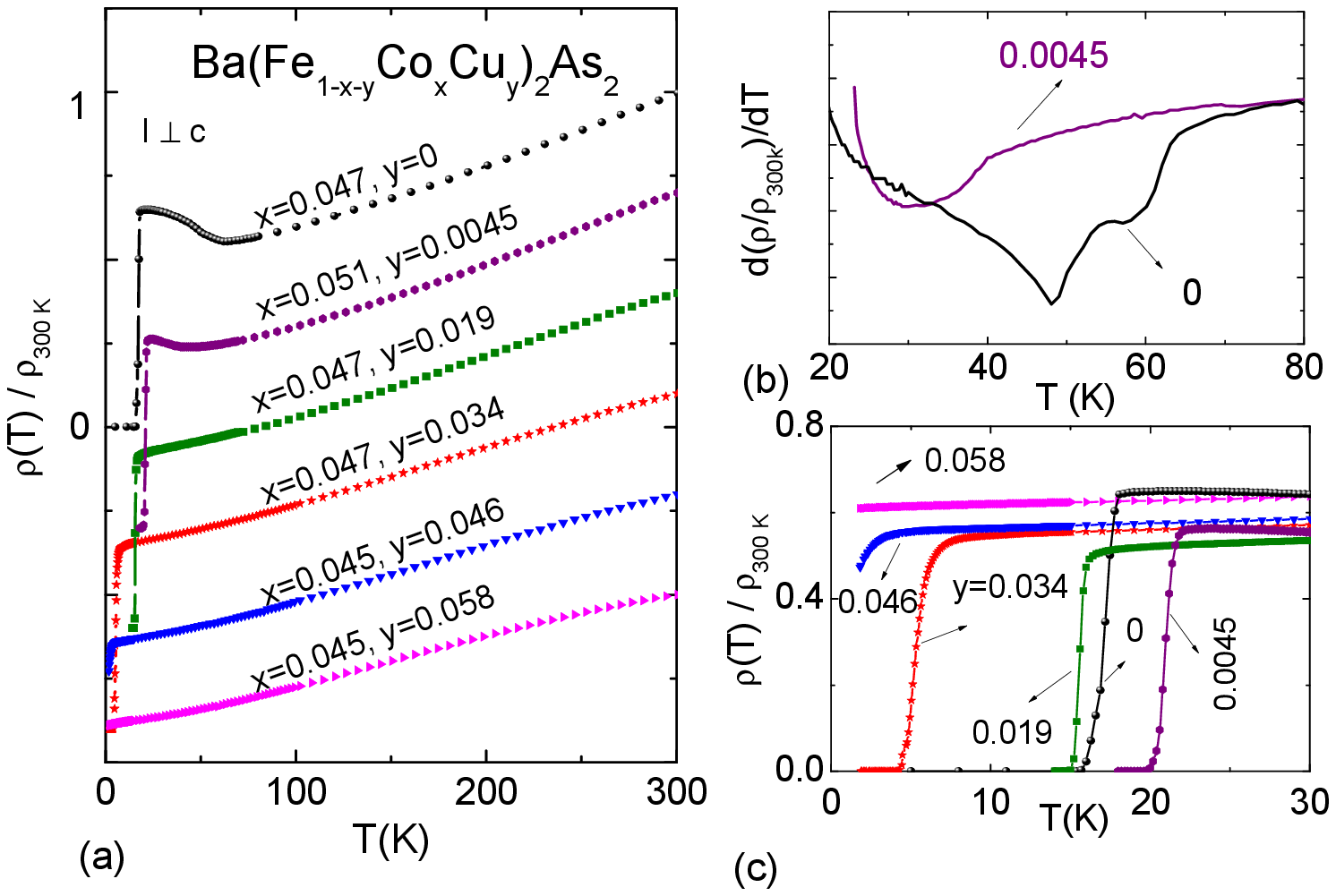,width=5in}
\caption{The ${\rm Ba(Fe}_{1-x-y}{\rm Co}_x{\rm Cu}_y)_2{\rm As}_2$ ($x \sim 0.047$) series: (a) The temperature
dependent resistivity, normalized to the room temperature value. Each subsequent data set is shifted downward by 0.3 for clarity.
(b) $d(\rho(T)/\rho_{300K})/dT$ for $y$=0 and 0.0045.
(c) Enlarged low temperature $\rho(T)/\rho_{300K}$.
}
\label{rcocub}
\eef

Figure \ref{rcocub} (a) shows the normalized resistivity of this series over the whole temperature range.
From the resistivity data, we can see that
${\rm Ba(Fe}_{0.953}{\rm Co}_{0.047})_2{\rm As}_2$ is an underdoped compound with $T_s=59$ K, $T_m=48$ K and $T_c\sim 17$ K.
With $y$ = 0.0045 of Cu doping, $T_c$ increases to 20 K and the structural / magnetic phase transitions
are suppressed to such an extent that only a resistance minima is detected before superconductivity
truncates the rest of the low temperature resistivity data (Fig. 22 (a) and (b)). The superconductivity feature can be more clearly seen
in Fig. \ref{rcocub} (c). For $y$ = 0.019 of Cu doping $T_c$ decreases to 15 K, there is no
longer any sign of structural and magnetic
phase transitions, and the resistivity has a
roughly linear temperature dependence above $T_c$. With even higher Cu doping, $T_c$ is suppressed to about 5 K for $y$=0.034 Cu doping. For $y$=0.046
of Cu doping, no zero in resistivity was measured down to 1.8 K, although some
decrease in resistivity around 2 K can be seen, which might suggest the onset of the superconducting state.
For $y$=0.058 of Cu doping, there is no sign of a superconducting
state.

\bef
\psfig{file=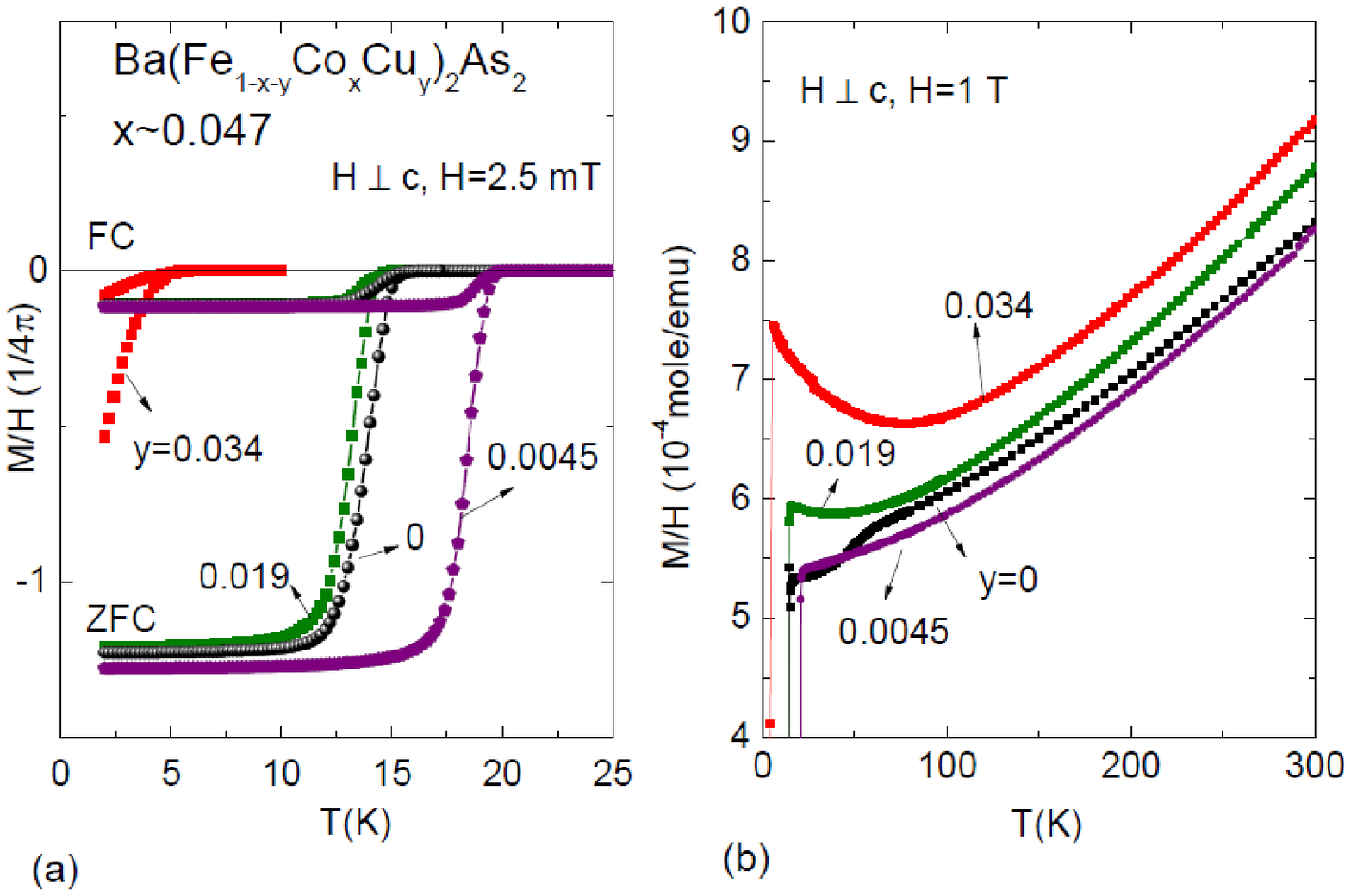,width=5in}
\caption{The ${\rm Ba(Fe}_{1-x-y}{\rm Co}_x{\rm Cu}_y)_2{\rm As}_2$ ($x \sim 0.047$) series:
 (a) Field-cooled (FC) and zero-field-cooled (ZFC) low field $M(T)/H$ data taken at 2.5 mT with $H\bot c$.
(b) $M(T)/H$ data taken at 1 T with $H\bot c$ for $0\leq y \leq0.034$.
}
\label{mcocub}
\eef

Figure \ref{mcocub} (a) shows the low field $M(T)/H$ data for this series taken at 2.5 mT with H perpendicular to the $c$ axis.
In FC measurements, the diamagnetic signal of the same magnitude
found for ${\rm Ba(Fe}_{1-x}{\rm Co}_x)_2{\rm As}_2$
suggests the same degree of
the bulk superconductivity in these samples as is found for the Co or Ni doped series.
The $T_c$ values inferred from the susceptibility data are consistent with the resistivity data.
Figure \ref{mcocub} (b) shows the temperature dependent $M(T)/H$ data taken at 1 T with H perpendicular to $c$ axis for $0.034\geq y \geq0$.
For $y=0$, a clear drop around 60 K can be seen in the susceptibility which is consistent with the structural / magnetic phase transitions
seen in the resistivity data. The second, lower temperature drop, around 20 K, is associated the superconductivity. With Cu doping $y\geq0.0045$,
no structural /
\bef
\psfig{file=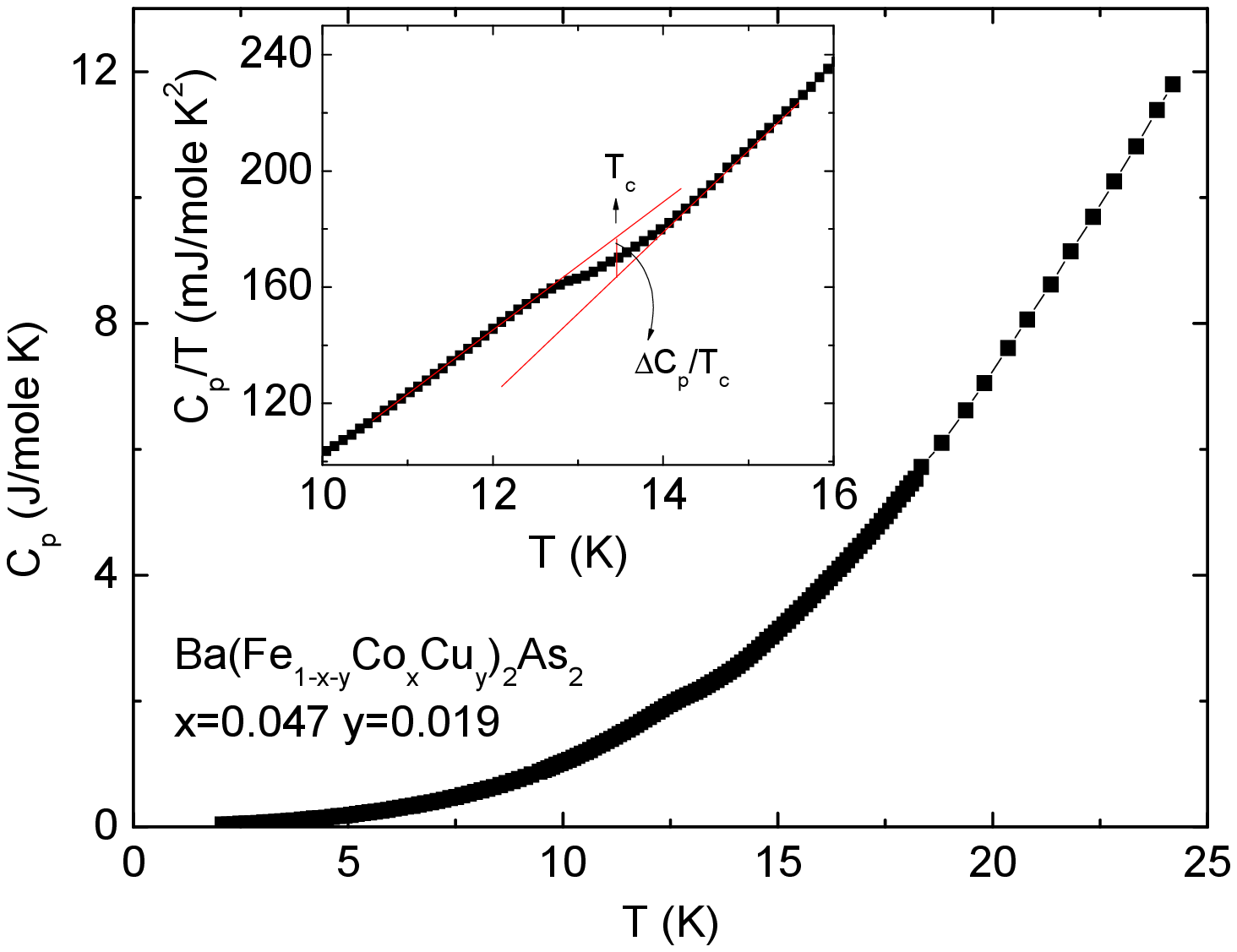,width=4in}
\caption{Temperature dependent heat capacity of ${\rm Ba(Fe}_{0.934}{\rm Co}_{0.047}{\rm Cu}_{0.019})_2{\rm As}_2$.
Inset: $C_p$/T vs. T near the superconducting transition with the estimated $\triangle C_p$ shown.}
\label{ckq378}
\eef
magnetic phase transitions feature can be seen although there is a minima, for $y=0.0045$, in the resistivity data.
The high temperature linear behavior in susceptibility is also observed in the
${\rm Ba(Fe}_{1-x-y}{\rm Co}_x{\rm Cu}_y)_2{\rm As}_2$ ($x \sim 0.047$) series.

Heat capacity data was collected for the first clearly overdoped member of this series:
${\rm Ba(Fe}_{0.934}{\rm Co}_{0.047}{\rm Cu}_{0.019})_2{\rm As}_2$, and is shown in Fig. 24. The heat capacity jump is consistent with the
bulk superconductivity in the sample. The inset shows the the enlarged $C_p/T$ vs. T data near $T_c$.
The inferred $\triangle C_p/T_c$ from "isoentropic" construction is 14 mJ/mole $K^2$ with $T_c$ equal to 13.4 K.
These vales also fall onto the $log(\triangle C_p/T_c)$ vs. $logT$ plot shown in reference \cite{sergey}.

Table \ref{tablecocub} summarizes these data and Fig. \ref{phasecocub} is a temperature-Cu doping concentration ($T-y$) phase diagram. It is worth
noting from Fig. \ref{phasecocub}
that with the addition of Cu in ${\rm Ba(Fe}_{0.953}{\rm Co}_{0.047})_2{\rm As}_2$,
$T_c$ does not decrease but rather increasing to $\sim 20 K$ at $y=0.0045$, and
probably has a higher value of $T_c$ for slightly higher $y$-values,
\bet
\begin{tabular}{c | c | c | c c c c|  c | c}
   \hline
   \hline

    dopant & $x$ & $y$  &\multicolumn{4}{|c|}{$\rho$}& \multicolumn{1}{|c|}{$M$}& \multicolumn{1}{c}{$C$}\\
    \hline
   Cu / Co& ~&~                      & {$T_s$}& {$T_m$}& {$T_c^{onset}$}& {$T_c^{offset}$} & {$T_c$} & {$T_c$}\\

    ~&      0.047 &0          &59     & 48       & 17.8  &16.5                                   &15.9  &      \\

  ~&        0.051 &0.0045      &$40^*$       &        & 21.5   & 20.4                            & 20.1 &      \\

   ~ &   0.047 &0.019       &      &          &15.9 &15.2                                 & 14.8 &13.5    \\

   ~       &  0.047 &0.034      &         &      & 6  &4.6                                    &5.7   &   \\

     ~     &   0.045 &0.046      &           &        &0     &0                               &  &  \\
     \hline
   \hline
  \end{tabular}
\caption{Summary of $T_s$, $T_m$ and $T_c$ from resistivity, magnetization and specific heat measurements for the
${\rm Ba(Fe}_{1-x-y}{\rm Co}_x{\rm Cu}_y)_2{\rm As}_2$ ($x \sim 0.047$) series.}
  \label{tablecocub}
\eet
\bef
\psfig{file=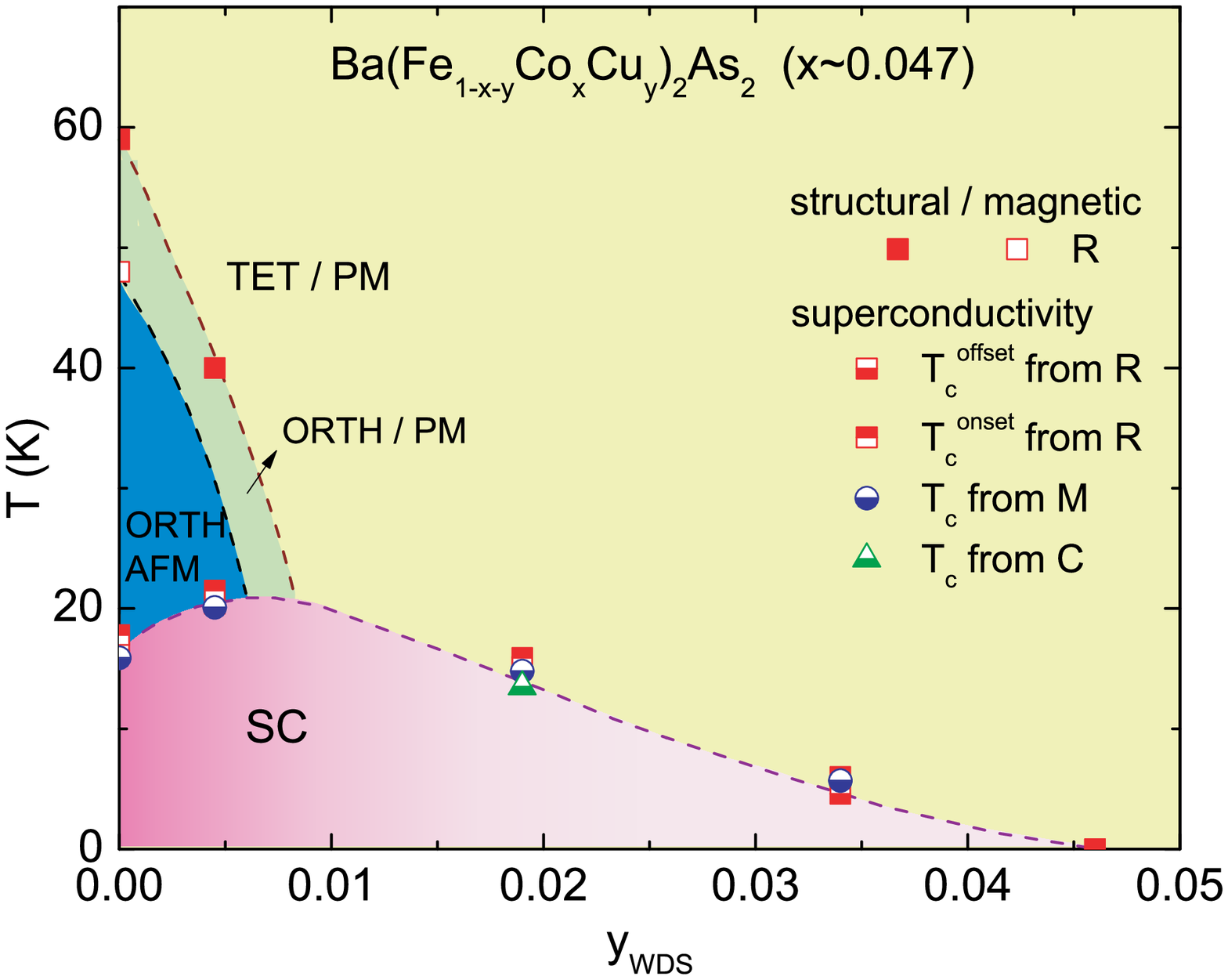,width=4in}
\caption{$T - y$ phase diagram of
${\rm Ba(Fe}_{1-x-y}{\rm Co}_x{\rm Cu}_y)_2{\rm As}_2$ ($x \sim 0.047$) single crystals.
The shading in the superconducting dome implies the existence of a crossover
from tetragonal / paramagnetic phase to orthorhombic / antiferromagnetic phase,
as used in Fig. \ref{phaseni}. Note: Given the rapid loss of features associated
with the antiferromagnetic transition, the AFM phase line is speculative.
}\label{phasecocub}
\eef
and then decreases to $\sim 15 K$ at $y=$ 0.019. These data, along with
the other Co / Cu doping series discussed in the previous section,
 clearly indicate that superconductivity can be induced and stabilized to relatively high $T_c$ values by Cu doping under well defined circumstances.

\section{discussion}

As we can see, in each series, good agreement in critical temperatures obtained from the
 resistivity, magnetization and heat capacity measurements has been observed.
The composite $T-x$ phase diagram, shown in Fig. \ref{3dphase} (a) highlights the similarities and differences between the various
${\rm Ba(Fe}_{1-x}{\rm TM}_x)_2{\rm As}_2$ series. For this diagram, $x$ was the total amount of TM dopants: e.g. for
${\rm Ba(Fe}_{0.953}{\rm Co}_{0.021}{\rm Co}_{0.026})_2{\rm As}_2$, $x$ would be 0.047. Figure \ref{3dphase} (a)
 is similar to the one shown in reference \cite{threedoping}, but it presents
a fuller Co and Cu doping data set as well as multiple Co / Cu doping data sets.
\bef
\psfig{file=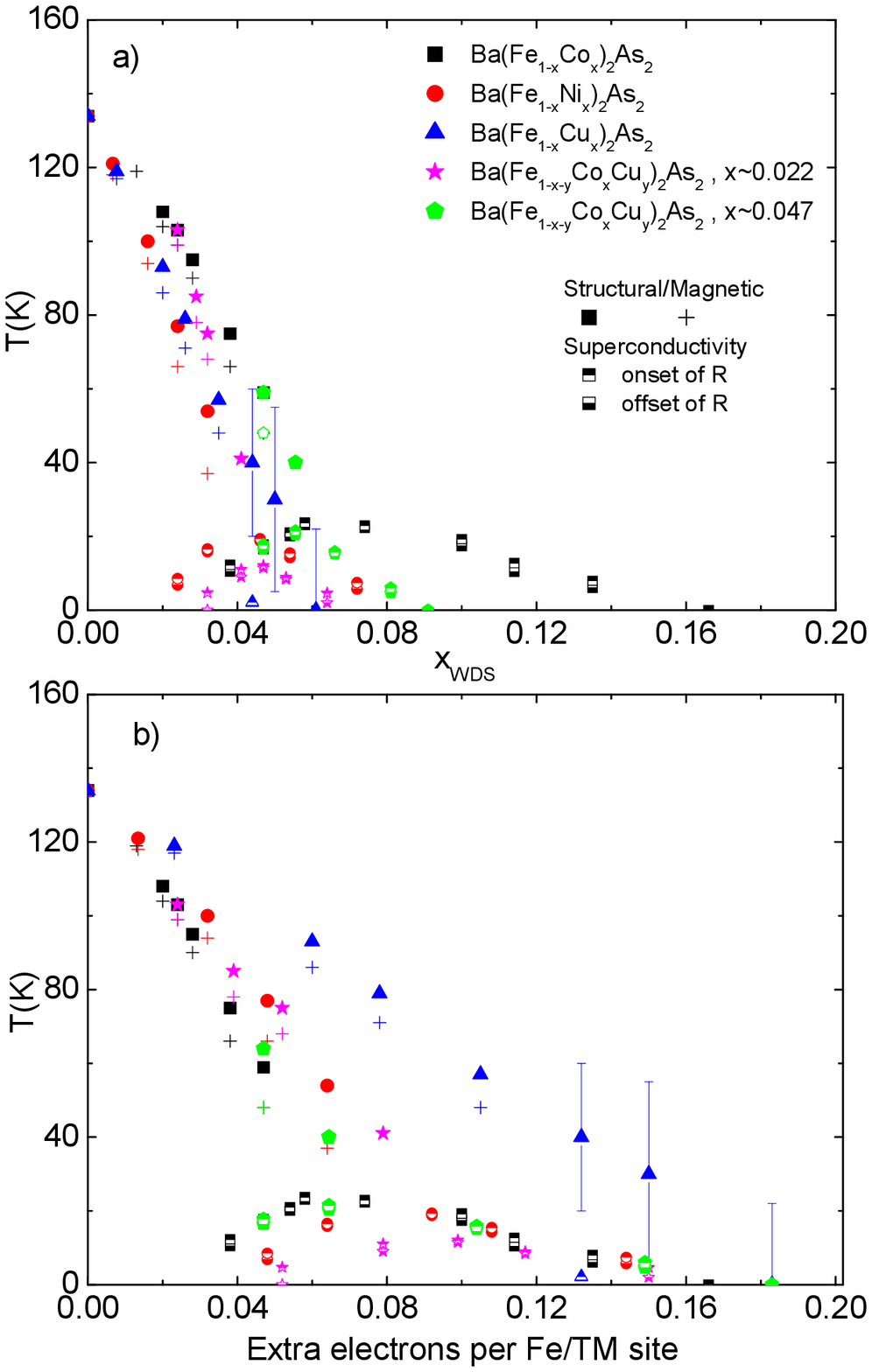,width=4in}
\caption{(a)$T-x$ phase diagrams for ${\rm Ba(Fe}_{1-x}{\rm TM}_x)_2{\rm As}_2$ (TM=Co, Ni, Cu, Co / Cu).
(a)$T-e$ phase diagrams for ${\rm Ba(Fe}_{1-x}{\rm TM}_x)_2{\rm As}_2$ (TM=Co, Ni, Cu, Co / Cu).
}
\label{3dphase}
\eef

The suppression rates of the upper phase transitions for
all these different series appear to depend on $x$, the number of TM substituted for Fe in a roughly similar manner
which appears to be inconsistent with a simple model of "nesting" induced
magnetism in these materials.
However the extent of the superconducting dome is not well described
by this parameterization. The ${\rm Ba(Fe}_{1-x}{\rm Co}_x)_2{\rm As}_2$ series
has the widest superconducting dome, ranging from $x \sim 0.03$ to 0.166.
${\rm Ba(Fe}_{1-x-y}{\rm Co}_x{\rm Cu}_y)_2{\rm As}_2$ ($x \sim 0.047$) has a dome extending to $x_{total} \sim 0.092$.
${\rm Ba(Fe}_{1-x}{\rm Ni}_x)_2{\rm As}_2$ ranks third with the dome starting at $x\sim 0.02$ and ending at $x\sim 0.075$.
The ${\rm Ba(Fe}_{1-x-y}{\rm Co}_x{\rm Cu}_y)_2{\rm As}_2$ ($x \sim 0.022$) series has an even narrower superconducting dome,
ranging from $x_{total}=x+y \sim 0.032$ to $x_{total}=x+y \sim 0.065$. The
${\rm Ba(Fe}_{1-x}{\rm Cu}_x)_2{\rm As}_2$ series has one superconducting point around $x=0.044$.

A closely related parameter, the extra
electrons added, $e$, can be inferred and the temperature-extra electrons phase diagram ($T-e$) can be constructed.
In this parameterization, a Co
dopant introduces one extra electron, a Ni dopant brings two extra electrons and a
Cu dopant adds three extra electrons. This leads to extra electron counts corresponding to
$x$ for Co doping, $2x$ for Ni doping, $3x$ for Cu doping, $x+3y$ for Co / Cu mixed doping. This
parameterization is consistent with our Hall resistivity and Seebeck coefficient measurements \cite{mun}.
This extra electron ($e$) parameterization is consistent with a simple "rigid band"
assumption for band filling, but is also consistent with recent proposals based on a density functional calculation that
the extra electrons are all localized around the dopant atoms \cite{sawatzky}, at its heart, the
extra electron parameterization simply assumes that one Ni atom has twice the effect of one Co atom and one Cu atom
has three times the effect of one Co atom.
Based on this parameterization, the $T-e$ phase diagrams are presented in Fig. \ref{3dphase} (b).
 As we can see, the superconductivity domes,
 especially on the overdoped side, are much better scaled by
this parameter.

A $T-e$ phase diagram similar to Fig. \ref{3dphase} (b) has already been mapped out in our earlier work
\cite {threedoping}. Via the fact that the structural, magnetic phase transitions (the superconducting domes) appear to be
parameterized by the doping level ( the number of additional electrons) respectively,
we suggested that superconductivity can be stabilized over a limited, and well delineated, range of $e$-values
when the structural and magnetic phase transitions are adequately suppressed.
For example, the data from the ${\rm
Ba(Fe}_{1-x}{\rm Cu}_x)_2{\rm As}_2$ series clearly demonstrate
that, if by the time the
structural / antiferromagnetic phase transitions are suppressed enough,
too many extra electrons have been added, the $e$-filling window for superconductivity can be missed.
On the other hand, if we adjust the position of the upper phase line in the $T-e$ phase diagram by judicious doping,
so that it does not miss the supercoducting window, superconductivity can occur.

Another way of seeing the different dependence of $T_s$ / $T_m$ and $T_c$ is to note that the
maximum $T_c$ value for a given doping series occurs where the extrapolated
$T_s$ / $T_m$ line hits the superconducting dome. When the data is plotted in a $T-e$ phase
diagram, it becomes clear that this point is where the $T_c-e$ data join the universal dome on the overdoped side.
By choosing the doping carefully, we can adjust the slope of $T_s(e)$ / $T_m(e)$ and to some extent
control where $T_c^{max}$ is. This is demonstrated by the Ba(Fe$_{1-x-y}$Co$_x$Cu$_y$)$_2$As$_2$ series:
by progressing from $x=0$ to $x=0.022$ to $x=0.047$, the $T_s$ / $T_m$ line acquires
a larger slope and $T_c^{max}$ increases.

The idea that the lower e-value extent of the superconducting dome is determined
by the rate of suppression of the $T_s$ / $T_m$ line carries with it the implication that
if this line could be suppressed even more rapidly, as a function of $e$, then $T_c^{max}$
could achieve even higher values. Unfortunately with $3d$- or even $4d$-transition metal
doping \cite{threedoping, ni4d}, Co and Rh have already offered the most efficient rate ($x:e = 1:1$). On the other hand $T_s$ / $T_m$
can be suppressed without any doping at all by the application of pressure.
Recent pressure measurements of $T-P$ phase diagrams for pure and Co-doped BaFe$_2$As$_2$ \cite{estell1, estell2}
show that indeed for pure and underdoped members of the Ba(Fe$_{1-x}$Co$_x$)$_2$As$_2$ series $T_c$ can be increased
significantly by suppressing $T_s$ / $T_m$ with pressure whereas over doped members of the series manifest
little or no increase in $T_c$ with pressure. Figure \ref{pressure1} summarizes the effects of pressure as well
as our $3d$ and $4d$ doping in the BaFe$_2$As$_2$ series. $T_c^{max}$ is extracted from the $T-P$ phase diagrams for
${\rm Ba(Fe}_{1-x}{\rm Co}_x)_2{\rm As}_2$ \cite{estell2}
and is selected as the highest $T_c$ value measured for a given $x$ under pressure. As we can see,
whereas $T_c^{max}$ differs only slightly from the $T_c$
values found at ambient pressure for the overdoped side of the superconducting dome, it continues
to rise for lower $x$ values, showing how large $T_c$ can be if $T_s$ / $T_m$ can be suppressed for lower $e$-values.
 These data  \cite{threedoping, ni4d} further emphasize that the two necessary, but not individually sufficient, conditions for
 superconductivity in this series seem to apply to different halves of the superconducting region:
 for the underdoped side of the dome, suppression of $T_s$ / $T_m$ is vital for superconductivity and for the overdoped side of
 the dome the value (and extent) of $T_c$ is defined by the value of $e$.

Figure \ref{pressure1} brings up a final important point: whereas for electron doping via TM substitution
in BaFe$_2$As$_2$, we appear to have a well defined pair of necessary, but not individually sufficient,
conditions for superconductivity, it should be born in mind that it is clear that the BaFe$_2$As$_2$
system can be tuned by other means. As clearly demonstrated pressure can tune $T_s$ / $T_m$ and $T_c$ and
produce $T-P$ phase diagrams that are topologically similar to the $T-x$ and $T-e$ phase diagrams we present here.
In addition P-doping on the As site and Ru doping on the Fe site are nominally isoelectronic dopings that
can also produce similar changes, albeit, at least in the case of Ru-doping for almost an order of magnitude higher doping levels \cite{rudoping, rudoping1}.
In all of these cases, either by electron doping on the TM site or by physical or "chemical" pressure it is likely that key features
in the band structure are being changed in some systematic manner. The challenge is to determine what that manner is.

\bef
\psfig{file=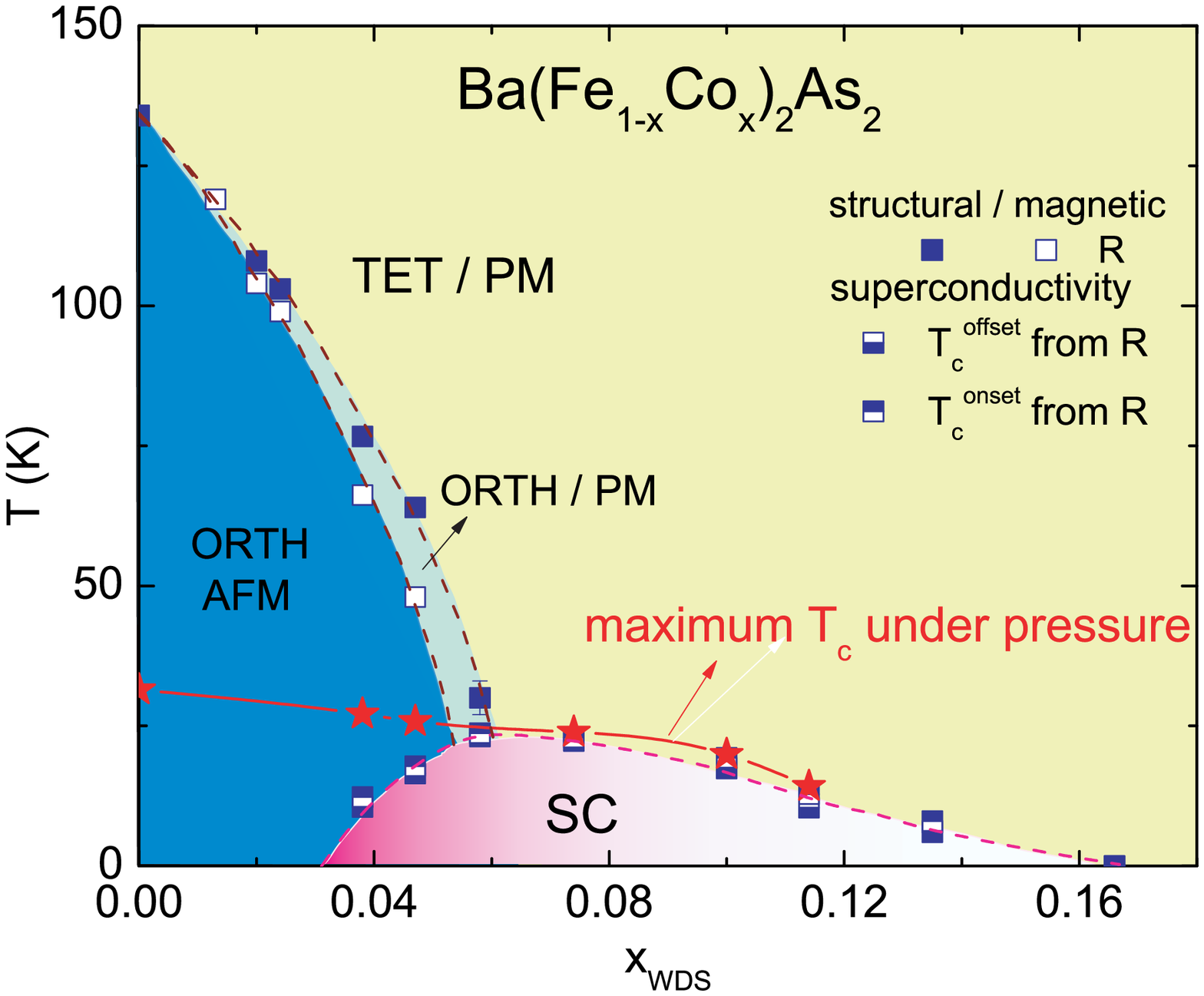,width=4in}
\caption{The comparison of the effects of chemical doping (\cite{more}) and application of pressure \cite{estell2} for
the Ba(Fe$_{1-x}$Co$_x$)$_2$As$_2$ series.
}
\label{pressure1}
\eef

\section{Conclusion}

Microscopic, structural, transport and thermodynamic
measurements have been performed on Ni-doped, Cu-doped as well as Co / Cu mixture-doped ${\rm BaFe}_2{\rm As}_2$ single crystals. Detailed
temperature-doping level ($T-x$) and temperature-extra electrons ($T-e$) phase diagrams have been mapped out for all these series. It was found
 the structural / magnetic phase transitions in pure ${\rm BaFe}_2{\rm As}_2$ at 134 K are monotonically suppressed in a similar
manner by these dopants. Superconductivity up to 19 K, 12K and 20 K can be stabilized in a dome-like region in the phase diagram for
Ni-doped, Co $_{\sim0.22}$ Cu-doped and Co $_{\sim0.47}$ Cu-doped series respectively while it is \emph{very} limited in Cu-doped series with only one measured
concentration ($x=0.044$) showing zero in resistivity near 2 K.
The application of 33 T external magnetic field on the optimally Ni doped ${\rm BaFe}_2{\rm As}_2$ sample
suppresses the superconducting temperature down to
0.6 $T_c(0)$ when $H\bot c$ and 0.3 $T_c(0)$ when $H || c$, indicating a small anisotropy with $\gamma$
varying from 2 (far from $T_c$) to 3 (near to $T_c$). Quantitative analysis of the $T-x$
and $T-e$ phase diagrams of these series reveals that the maximum $T_c$ value for a series occurs close to where the extrapolated
$T_s$ / $T_m$ line intersects the superconducting dome and that the rate of the suppression of $T_s$ and $T_m$ is governed by $x$
whereas $e$ appears to parameterize the envelop of the superconducting dome. The comparison between the effects of
chemical doping and application of pressure for Ba(Fe$_{1-x}$Co$_x$)$_2$As$_2$ series further reveals that $T_c$ in the underdoped region is
controlled by the extent $T_s$ / $T_m$ are suppressed whereas it is defined by the $e$ value for the overdoped region. Therefore,
by choosing the combination of dopants are used we can adjust
the relative positions of the upper phase lines (structural and magnetic phase transitions) and the
superconducting dome to control the superconductivity in electron-doped ${\rm BaFe}_2{\rm As}_2$.


\section*{Acknowledgments}
We would like to thank M. E. Tillman for the assistance in the high magnetic field $H_{c2}$ measurement, N. H. Sung for the help in
samples growth, E. D. Mun for the aid in the measurements. Work at the Ames Laboratory was supported by
the Department of Energy, Basic Energy Sciences under Contract No.
DE-AC02-07CH11358.



\begin{thebibliography}{99}
\bibitem{more}
New criterion to infer $T_s$ has been applied to the data presented in \cite{nico};
 three new batches of samples with $x=0.024, 0.135$ and 0.166 have been grown and characterized.
\end{thebibliography}
\end{document}